\documentclass[12pt,JSTAT]{iopart}
\usepackage{graphicx}
\usepackage{amssymb}
\usepackage{colortbl}
\usepackage{pstricks}
\usepackage{bm}
\usepackage{latexsym}
\usepackage[english]{babel}
\usepackage[latin1]{inputenc}

\begin{document}

\title{Finite size analysis of a double crossover in transitional wall turbulence}


\author{Joran Rolland$^1$}

\address{ Institut PPRIME, UPR 3346, Chasseneuil du Poitou, France.}
\ead{jrolland@phare.normalesup.org}

\begin{abstract}
This article presents the finite size analysis of two consecutive crossovers leading laminar-turbulent bands to uniform wall turbulence
 in transitional plane Couette flow. Direct numerical simulations and low order modeling simulations of the flow are performed.
 The kinetic energy $E$ of the turbulent flow and the order parameter $M$, a measure of the spatially organised modulation of turbulence,
 are sampled. These two quantities are processed in view analytical results from the phenomenology of phase transitions. The first crossover concerns the loss
 of spatial organisation of turbulence in the flow. In the band phase, the order parameter $M$ decreases continuously with the Reynolds number $R$ toward a small value,
 while its response function $\chi_M$ displays a maximum at the crossover. The increase of the maximum of the response function  $\max \chi_M$ with domain size is consistent with a polynomial law $\max \chi_M \propto (L_xL_z)^{\frac{\tilde{\nu}}{\mu}}$, $\tilde{\nu}/\mu\simeq 1$.
 A first critical Reynolds number $R_{c,1}$ can be defined as the value at which the maximum of the response function is reached. In the uniform phase,
 the order parameter $M$ and its variance $\sigma$ decrease toward zero following mean field scalings $M,\sigma \propto 1/\sqrt{L_xL_z(R-R_c)}$ as $R$ is increased.
 The kinetic energy $E$ is an affine function of $R$  except in a small range where a sharp increase is detected, which corresponds to the second crossover.
 A second critical Reynolds number $R_{c,2}$ can be defined as the center of this sharp increase range. In this range, spatial and temporal coexistence of
 the uniform turbulence phase and laminar-turbulent bands phase is observed. This sharp increase is concomitant with a maximum of the response function of the
 kinetic energy. The finite size analysis reveals that the jump does not steepen and that the maximum of response function of $E$ saturates as size is increased.
 The first crossover is formally identical to a critical phenomenon in condensed matter. The second crossover is in agreement with a first order phase
 transition smeared by finite noise. The analytical analysis of this phenomenon assuming a non interacting gas of fronts between domains of
 the two phases provides a scaling of the response function consistent with that of $E$. In our context, this amounts to the statistics of the grain boundaries between domains of banded turbulence and uniform turbulence. We eventually discuss how this formalism could explain a breakdown of orientation order of the bands, with grain boundary between domains of different orientations. This breakdown may occur in extremely large size domains and could affect the order parameter and its response function.

\end{abstract}

\noindent{\it transition to turbulence, wall turbulence, bifurcation, phase transitions}

\submitto{J. Stat. Mech.}

\maketitle

\section{Introduction}\label{intro}


The formation of large scale coherent flows is common in developed or transitional turbulence.
The type of organisation is not unique and one can often observe crossovers between large scale flow topologies when a control parameter is changed.
These crossovers are often accompanied by very large fluctuations of a relevant scalar observable and/or multistability
between large scale flow configurations. This is for instance the case in two dimensional turbulence \cite{fetal,BSsns},
in Von K\'arm\'an flow \cite{karman} or in ultimate Taylor Couette turbulence \cite{multi_tc}.
A key point in the study of these crossovers
is the use of statistical physics methods to predict the structure of the large
scale flow for each value of the control parameters \cite{fetal,BSsns}, infer the mean time spent in each of the flow configurations,
or make sense of the fluctuation maxima of the most relevant observables \cite{karman}.
In each of these cases, the parallel to a phase transition (continuous or discontinuous, at or outside equilibrium) is drawn,
although it is not possible to take a thermodynamic limit in such systems.

\begin{figure}
\centerline{\textbf{(a)}\includegraphics[width=7cm]{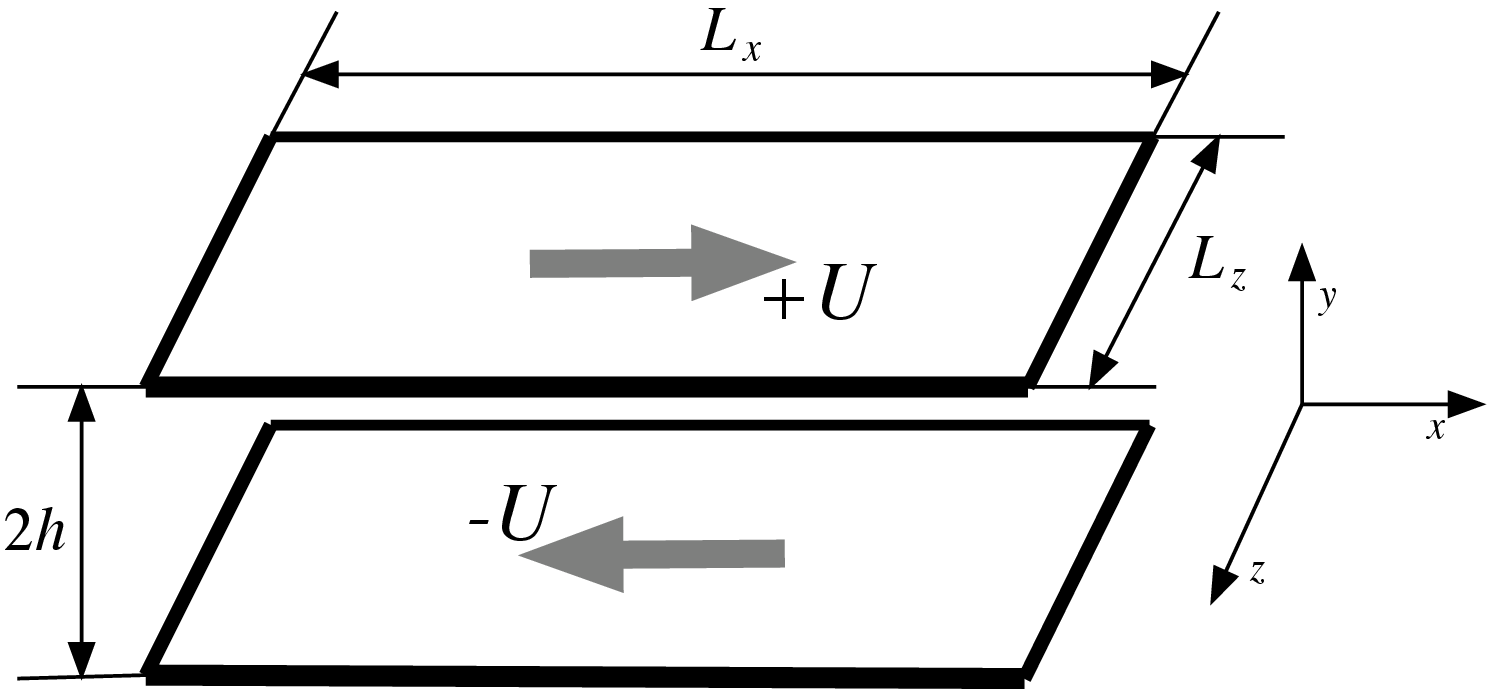}\textbf{(b)}\hspace{1mm}\includegraphics[width=5.5cm,clip]{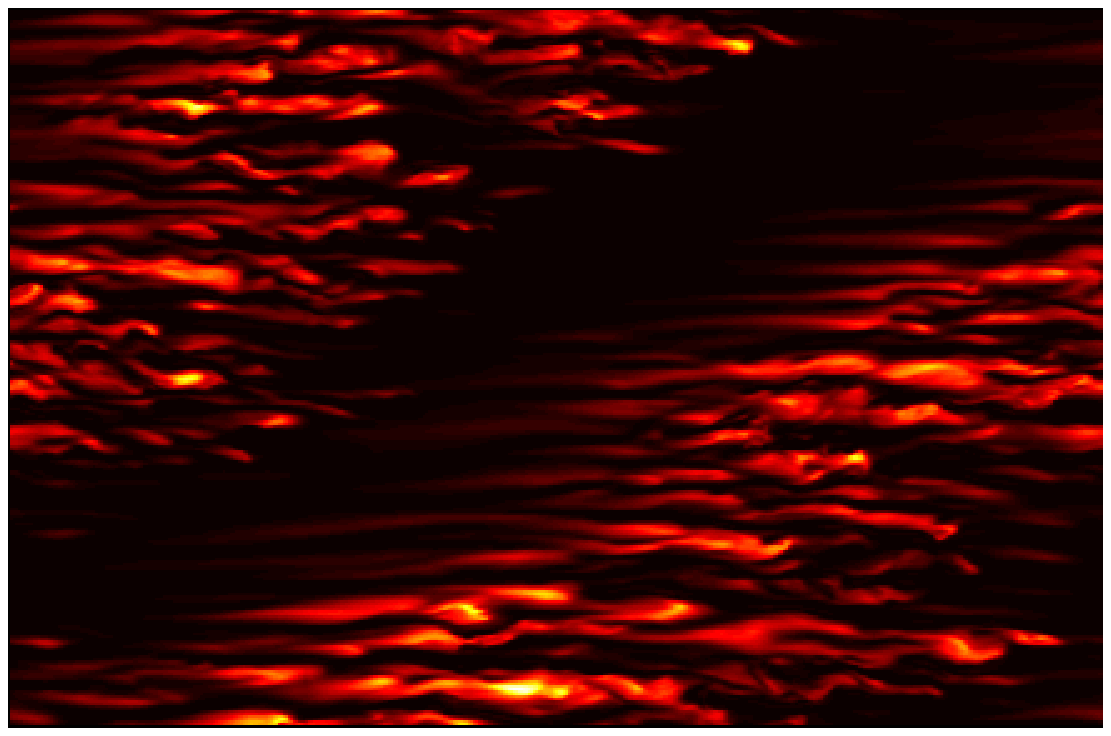}}
\caption{(a): Sketch of the configuration of plane Couette flow, indicating the movement of the walls, the axes and the dimensions of the system. (b): Illustration of a laminar-turbulent band with colour levels of turbulent kinetic energy in a $x- z$ plane.}
\label{figint}
\end{figure}

Another type of crossover for which these concepts may actually be used entirely is the late afternoon transition of the planetary boundary layer \cite{PBL,pbl,ch}.
In that case, the radiative heating decreases at the end of the day, so that the density stratification turns from unstable to stable.
Wall and convective turbulence are strongly reduced in amplitude and the flow is reported to be very \emph{intermittent}.
It has been shown that the core mechanisms and phenomenology of this crossover are those of neutrally buoyant
wall turbulence. In that case the flow is controlled by the Reynolds number, the ratio of advection over viscosity (Fig.~\ref{figint} (a)). For Couette flows one precisely has $R=Uh/\nu$ with $U$  the wall velocity, $h$ the half width of the channel and $\nu$ the kinematic viscosity.
 In these flows, when the Reynolds number is decreased,
one first observes a transition between uniform wall turbulence and laminar turbulent coexistence \cite{PBL,BDS}. This transition occurs
in a range of Reynolds numbers around $R_{\rm t}$, for which several  definitions exist \cite{BT11,RM10_1,prigent02}. With all these definitions, one has $R_{\rm t} \in [390;440]$ for experiments or DNS of Couette flows.
This coexistence is often spatially organised, like the spiral turbulence of Taylor--Couette flow \cite{cva}:
one observes laminar-turbulent bands. The bands are oblique with respect to the streamwise direction. This is typical of transitional wall flows which can extend in two directions of space (see Figure~\ref{figint} (b)).
When the Reynolds number is further decreased, one observes a second transition to a regime where turbulence cannot sustain itself at a Reynolds number $R_{\rm g}\simeq 325 \pm 5< R_{\rm t}$ \cite{natphyscoup,bpre}.


  The second transition at $R_{\rm g}$ has led to the introduction of spatio-temporal intermittency to describe transitional turbulence \cite{pphysd,M90}. This is a first instance of a transition where statistical physics descriptions have been successful. Over the last three decades many  experimental, numerical and modeling studies have investigated the phase transition associated with spatiotemporal intermittency (see \cite{pre_sano_09} for a synthesis). The recent access to increased computational power and larger experimental facilities has brought back a lot of attention on this phenomenon in wall flows \cite{natphyscoup,bpre}. The first transition, from coexistence to uniform turbulence, can possibly be described as a phase transition,
  but has received comparably little attention \cite{RM10_1}. It shows a crossover between the laminar turbulent coexistence and uniformly
  turbulent flow in a setting where one can easily take a thermodynamic limit. However, the picture is more complex since several scenarii can be proposed,
  which strongly depend on the number of dimensions in which the flow can extend. If the flow can extend in two dimensions, like in Couette flows,
  turbulence is structured in bands which correspond to a sinusoidal modulation of the amplitude of turbulence \cite{RM10_1,BT11,prigent02,PBL,BDS}.
  The amplitude of modulation of turbulence decreases continuously as the Reynolds number is increased \cite{RM10_1,BT11,prigent02}.
  Near the disappearance of the coexistence, this amplitude undergoes very strong fluctuations mostly linked to switching of orientations
  of the laminar turbulent bands. The combination of the decrease of amplitude of modulation of turbulence and the fluctuation crisis of said
  amplitude are reminiscent of a critical phenomenon. The picture is made more complex by the fact that the kinetic energy of the flow can also
  be used as a measure of this crossover. In that case, it shows a change of regime at a slightly higher Reynolds number with strong fluctuations.
  This time, they are associated to the time and space coexistence of the two phases: laminar turbulent bands and uniform turbulence \cite{RM10_1,BT11}.
  This phenomenology is then more reminiscent of a first order phase transition. In ``two dimensional flows'',
  a combination of two successive phase transitions for two different order parameters is a possible scenario,
  comparable to the melting of a two dimensional system of hard disks \cite{hd,melting_hd}. In ``one dimensional'' flows like Hagen-Poiseuille flow,
  only the second crossover takes place \cite{bpre}. Thus, additional care must now be taken in order to identify the transition types.
  An approach where one can identify undoubtedly the scenarii, while precisely characterising the crossover in the highly fluctuating  context of $R\simeq R_{\rm t}$,
  should be proposed.


  The crossover at $R_{\rm t}$ has first been studied from a dynamical systems point of view. Ginzburg--Landau models from pattern formations have first been introduced to describe the disappearance of the modulation of turbulence at $R_{\rm t}$ \cite{prigent02}. Due to the fluctuating nature of transitional turbulence \emph{ad hoc} noise had been added to these empirical models. The notion that a change of regime, a bifurcation, was still occurring was reconciled with this fluctuating case by the use of pdf bifurcations, where fitting parameter of the logarithm of sampled pdf are seen to change sign. This notion of bifurcation is actually well defined mathematically for stochastic dynamical systems as phenomenological bifurcations (noted $\mathbb{P}$-bifurcation, see \cite{HD} \S~2,~4, \cite{LA}). However, this approach is very dependent of the choice of fitting function, while more univocal, dynamical notions of bifurcations in stochastic dynamical systems\footnote{(dynamical) $\mathbb{D}$-bifurcations, introduced in non-autonomous dynamical systems \cite{LA}} are actually particularly heavy to implement in our turbulent flows. This further encourages us to partially leave dynamical systems and use methods derived from statistical physics. In the case of phase transitions in condensed matter, the systems are already at thermodynamic limit,
  so that identifying the transition type, and measuring all critical exponents if the transition is a critical phenomenon,
  is relatively straightforward.
  However, in numerical simulations of phase transitions in statistical physics, the systems
  (of dimension $d$) always have a finite size $L$ and finite volume $L^d$ \cite{BL}.
   In the case of critical phenomena, this means that
  the order parameter does not strictly go to zero in the disordered phase, while there is no strict discontinuity at the transition
  if it is of the first order. The response function, $\chi=\sqrt{L^d}\times$ fluctuations of the order parameter, does not diverge.
  Finally, the value of the control parameter at which the finite size crossover occurs depends on size. This makes the identification
  of scenarii difficult: one can mistake a first order transition with a second order transition. Indeed, in both cases the response function has a peak.
  One can even wrongly identify an ordered phase when no order or quasilong range order exist in the thermodynamic limit.
  The cost of estimating precisely the critical exponents, if the transition is of second order, can be tremendous. Hopefully,
  an approach at studying numerically phase transitions which circumvents these issues has been used for a long time.
  It is termed finite size analysis and takes advantage of the fact that the maxima of response functions, control parameters
  at which the transition occurs \emph{etc}. follow specific power law scalings in size which depend on the order of the transition \cite{BL}.
  In second order phase transitions, which will be central in the discussion of this text, the maximum of the response function $\chi^m(L)$ at size $L$ and the control parameter $\mathcal{R}(L)$ value at which this maximum is reached follow scalings of the form
  \begin{equation}
  \chi^m\propto L^{\frac{\tilde{\nu}}{\mu}}\,,\, \mathcal{R}(L)\propto \mathcal{R}(\infty)+a L^{-\frac{1}{\mu}}\,, \label{FSA_i}
  \end{equation}
  where $\tilde{\nu}$ is the exponent describing the divergence of $\chi\propto|\mathcal{R}-\mathcal{R}(\infty)|^{-\tilde{\nu}}$
   in infinite size domains, and $\mu$ is the exponent describing the divergence of the correlation length $\zeta\propto|\mathcal{R}-\mathcal{R}(\infty)|^{-\mu}$
   in infinite size domains.
  This means that one can sample the order parameter in the $R_{\rm t}$ range, its response function \emph{etc}. in systems of not so large, increasing sizes
  $L\in[L_{\rm min}; L_{\rm max}]$, and identify the type of transitions and critical exponents (if relevant). This requires mostly
  \emph{one parameter fits} and has a degree of precision much higher than a direct estimation from a sampling at $L_{\rm max}$, for a comparable cost.
  Moreover, this approach can be transferred to percolation type transitions at $R_{\rm g}$, this has for instance been used to successfully compute
  \emph{all} exponents and relations between them in a non-linear optic system \cite{JTM}. We will therefore follow this point of view when sampling and processing our data.

The text is organised in the following manner. We first present the numerical
and processing methods in section~\ref{num}. We then present in more details the phenomenology of
the transitions in plane Couette flow in section~\ref{phe}. The quantitative results of the finite size analysis
of the crossovers and the  uniform turbulence phase are given in section~\ref{res}.
The results are eventually discussed in section~\ref{concl}. Additional analytical results used in the text are derived in~\ref{App} and~\ref{fstod}.

\section{Simulations of plane Couette flow}\label{num}

\subsection{Configuration and numerical method}

We consider transitional turbulence in plane Couette flow. Let us define the scales and the axes of our system.
Plane Couette flow is the flow between two parallel planes separated by a gap $2h$, moving at velocities $+U \vec e_x$ and $-U \vec e_x$
with a dynamical viscosity $\nu$ (Figure~\ref{figint} (a)). The streamwise direction is $\vec e_x$,
$\vec e_y$ denotes the wall normal direction and $\vec e_z$ is the spanwise direction.
We work with dimensionless equations: $U$ scales the velocities, $h$ scales the lengths and $h/U$ scales the durations.
We define the Reynolds number, the main dimensionless control parameter, as  $R\equiv hU/\nu$.
The streamwise and spanwise sizes of the periodical domain, $L_x$ and $L_z$ (already rescaled by $h$), will be
two other fundamental control parameters of the flow. When considering the rescaled incompressible Navier--Stokes equations,
 the laminar baseflow is $\vec v=y\vec e_x$. The departure to the laminar baseflow is denoted by $\vec u$,
so that one has the full velocity field $\vec v=y \vec e_x+\vec u$.

We perform a numerical study. We integrate in time $\vec u$ the departure to the laminar baseflow. For this matter, the velocity field $\vec u$ is
projected on $N_x$ and $N_z$ dealiased Fourier modes for spatial dependence in $x$ and $z$ and on $N_y$ Chebyshev
modes for wall normal spatial dependence $y$. The numerical integration is performed using the code {\sc Channelflow}
by J. Gibson \cite{gibs}. We will perform classical Direct Numerical Simulations with in plane resolution
$N_x/L_x=N_z/L_z=8/3$ Fourier modes and wall normal resolution $N_y=27$ for illustrative purposes.
The complete physical study will be performed using a lower order modeling procedure with $N_y=15$ Chebyshev modes,
$N_z/L_z=2$ spanwise Fourier modes and $N_x/L_x=1$ streamwise Fourier modes. Our focus in placed on a
thorough physical analysis rather than on a match of classical diagnostics between numerics and experiment.
Indeed, the physical study can be thoroughly performed at a much lower cost: the features of the transition,
the universal behaviours (scaling law of amplitude of modulation, \emph{etc}.) are unchanged \cite{MR10,RM10_1,RM10_2,CTB}.
The trade-off is a change of non universal quantities: for instance transition thresholds are decreased
from $R_{\rm g}\simeq325$ and $R_{\rm t}\simeq400$ to $R_{\rm g}\simeq 275$ and $R_{\rm t}\simeq345\pm 10$
(the exact value of each threshold depends on the definition) \cite{MR10}. Other features of transitional wall flows,
such as traveling waves or exact coherent solutions, can be dramatically impacted by resolution (or preimposed symmetry).
For instance, they can disappear when $N_y$ is increased \cite{MR10}. Note also that if resolution is further decreased
(either below $N_y\le11$ for a discretisation with Chebyshev polynomials \cite{MR10} or with $N_y=3$
with a basis incorporating boundary conditions \cite{M09}), part of the ordering of turbulence disappears:
oblique bands no longer exist. In a way, the system jumped into another university class.

Initial conditions for the numerical integrations are generated in the following manner. We first generate a
random velocity field by drawing Fourier modes according to a Gaussian law whose variance decreases like a
powerlaw of the wavenumber. This random velocity field is then integrated for $500$ time units at $R=450$
(for the low order procedure) or $R=500$ for the DNS so that it reaches steady state homogeneous wall turbulence.
The flow can then be brought to a lower Reynolds number by quench or adiabatic decrease of $R$ for data sampling.
Further integration at this Reynolds number is performed until a statistically steady state is reached and sampling can begin. This typically requires that the
flow selects only one orientation of the bands in the whole domain. In the largest domains, the domain can accommodate bands of two or more wavelengths and angles.
Sampling is performed using an initial condition which contains the most probable wavelength and angle, at each given Reynolds number. This can be done by first preparing spatially organised
banded velocity fields which display each likely wavelength, for instance through quenches at different Reynolds number.
Each of these velocity fields is integrated in time at the Reynolds number of interest for duration of order of several
$\mathcal{O}(10^4)$. In simulations started with a band wavelength and angle which are not favored, a change of wavelength will occur
and all simulations will eventually display the same wavelength.  further integration  is performed to ensure that the flow indeed keeps the selected band wavelength.

Let us eventually present the dataset. For the studies presented in sections~\ref{phe} and~\ref{res},
we considered systems of sizes ranging from $L_x\times L_z=56\times 48$ to $L_x\times L_z=440\times 192$.
In the range around $R_{\rm t}$  the data was sampled at least at each values of $R$ (this range is a size of typically 15 Reynolds number units, and every half Reynolds number unit at $R_{\rm t}$). Outside this range
the sampling was not so fine in Reynolds number: we typically have sampled time series every $10$ units of $R$. The necessary duration of
sampling is not necessarily long outside the $R_{\rm t}$ range, it is of order $\mathcal{O}\left(10^4 \right)$.
Inside the $R_{\rm t}$ range, we will meet strong fluctuations, so that longer datasets are needed for a good estimation.
Practice showed that at least one reversal of orientation or another strong fluctuation should be sampled for acceptable estimations.
 This means that the datasets at each Reynolds number and size have a duration of order $\mathcal{O}\left( 10^5 \right)$.
Similar sampling duration is also used when a change of wavelength has been detected: it ensures that leaving is far less probable than coming.

\subsection{Processing}\label{proc}

  Several spatially averaged diagnostics are used to describe the flow at any given time. They are regularly
used in numerical and experimental studies of transitional wall turbulence \cite{BT11,RM10_1,RM10_2,M11,MR10,epjb15,epje16}.
These diagnostics are averaged one way or another in order to characterise the steady state for given values of Reynolds number $R$,
sizes $L_x$ and $L_z$. The simplest one is termed the kinetic energy of the flow and is defined as
\begin{equation}
e(t)\equiv \frac{1}{2L_xL_z}\int_{x=0}^{L_x}\int_{z=0}^{L_z}\int_{y=-1}^1(u_x^2+u_y^2+u_z^2)\,{\rm d}x{\rm d}z{\rm d}y\,.
\end{equation}
It indicates us how much energy is contained in all the coherent structures of the flow. It is a function of time.
Let us set $t=0$ as the time where the flow has reached the neighbourhood of the statistically steady state. Data is sampled up to time $T$,
the average kinetic energy then reads
\begin{equation}
E\equiv \frac{1}{T}\int_{t=0}^Te(t)\,{\rm d}t\,.
\end{equation}
Note that $e(t)$ is a fluctuating quantity. We sampled pdf of $e$ during the interval $[0;T]$. In order to measure the amount of fluctuations, we can also compute the variance of $e$
\begin{equation}
\sigma_E\equiv\sqrt{\left(\frac{1}{T}\int_{t=0}^Te^2(t)\,{\rm d}t\right)-E^2}\,.
\end{equation}
In practice, in the study of phase transitions, the relevant, finite, quantity is the response function \cite{BL}.
We therefore define the response function of the kinetic energy $\chi_E\equiv \sqrt{L_xL_z}\sigma_E$.

We then define a turbulent fraction $f(t)$ (sometimes called an intermittency factor). Its computation from the velocity
field first requires the identification of turbulent and laminar zones \cite{RM10_1,pm11,M11}. We reuse an approach
which has already proven successful \cite{RM10_1}: the domain is divided in subdomains of size $l_x=2$, $l_z=2$, $l_y=1$.
Each subdomain corresponds either to $y<0$ or $y>0$. The square norm of the velocity $u_x^2+u_y^2+u_z^2$ is spatially
averaged in each subdomain. If this norm is larger than a given threshold $c=0.025$, the cell is considered turbulent,
if it is smaller than $c=0.025$ the cell is considered laminar. This yields a laminar/turbulent discriminated field
which gives us a three dimensional coarse grained view of the localisation of turbulence.
This laminar/turbulent discriminated field can be displayed in two dimensional colour maps in a $(\vec e_x,\vec e_z)$
(Figure~\ref{reen}). At each $x,z$ position, we set a different colour depending on whether we
find two laminar cells for $y<0$ and $y>0$ (black), two turbulent cells for $y>0$ and $y<0$ (white) or a turbulent
cell for $y>0$ and a laminar cell for $y<0$ (or vice versa, colour). This indicates us the shape and size of laminar
and turbulent domains, especially if they take a banded form (Figure~\ref{figint} (b)). At a given time, the turbulent fraction
$f$ is the number of turbulent cells divided by the total number of cells.

 A quantity like $E$ can be a relevant order parameter when studying a transition between a uniformly turbulent flow and
 any laminar-turbulent coexistence, regardless of its spatial organisation. However, it yields no indication as to a
 spatial organisation, in the form of bands for instance (Fig.~\ref{figint} (b), Fig.~\ref{reen}).
 For this, we sample the Fourier modes of the streamwise velocity field at wavenumbers $k_x=2\pi/\lambda_x$ and
 $k_z=2\pi/\lambda_z$, where $\lambda_x$ and $\lambda_z$ are the wavelengths of the bands. These wavelengths are first identified through visualisations of the velocity field (such as Fig.~\ref{figint} (b) and Fig.~\ref{reen}).  We then compute a pair of
order parameters for the spatial organisation of turbulence into bands $m_\pm(t)$ by averaging the Fourier component $|\hat{u}_x|^2$ over height
 \begin{equation}
 m_{\pm}(t)=\sqrt{\frac{1}{2}\int_{-1}^1|\hat{u}_x|^2(k_x,y,\pm k_z,t)\,{\rm d}y}\,.
 \end{equation}
For safety, time series of $m_\pm$ are systematically sampled at $k_x$ and $\pm k_z$, as well as at neighbouring wavenumbers: this helps confirm the relevance of the choice of $k_x$ and $\pm k_z$, since the corresponding Fourier mode is much larger than its neighbours.
This type of observable
 is commonplace in the study of laminar turbulent bands \cite{BT11,prigent02,RM10_1}.
There is a symmetry between $+$ and $-$ orientation which is broken most of the time when the system is in a
steady state (Figure~\ref{figint} (b), Figure~\ref{reen}). We take this fact into account in order
to compute the average, variance and response function of $m$. We have to be all the more careful that the flow can
experience global reversals of the orientation of the bands: both states can be visited with equal probability \cite{RM10_2,prigent02}. We follow the same line of thought as \cite{BT11} and use the method defined in \cite{RM10_1}
and sample joint probability density functions of $m_+$ and $m_-$: $\rho(m_+,m_-)$. The sampling of $m_\pm$ starts
at the same time $t=0$ as that of $e(t)$. The PDF is symmetrised along the $m_+=m_-$ axis in order to accelerate the
sampling \cite{RM10_1}. The probability density function is bimodal if $R\lesssim R_{\rm t}$: this is a
rephrasing of the equal probability of occurrence of both orientations. We can eventually define the order parameter $M$ as
\begin{equation}
M\equiv \int_{m_+,m_-<m_+}m_+\rho(m_+,m_-)\,{\rm d}m_+{\rm d}m_-\,.
\end{equation}
We can then compute the variance of this order parameter
\begin{equation}
\sigma_M\equiv \sqrt{\left(\int_{m_+,m_-<m_+}m_+^2\rho(m_+,m_-)\,{\rm d}m_+{\rm d}m_-\right)-M^2}\,.
\end{equation}
We also define the response function of this order parameter $\chi_M\equiv\sqrt{L_xL_z}\sigma_M$.

\section{Phenomenology}\label{phe}

In this section, we describe direct numerical simulations in one domain of relatively small size in the whole range of Reynolds numbers as well as
a low order modeling simulation in larger size domain showing a typical event. Both will serve as a point of comparison for the systematic study.


We first consider the averages and response functions sampled in direct numerical simulations.
In order to save computation power and present what we will systematically study with the low order modeling, in particular with the kinetic energy,
we will only consider one size: $L_x\times L_z=110\times 32$, containing one wavelength of the band, for the whole Reynolds number range.
On top of this, this section serves as the verification that there are only quantitative shifts in Reynolds number
values and some amplitudes between the DNS and the low order simulations.

\begin{figure}
\centerline{\includegraphics[width=6.5cm]{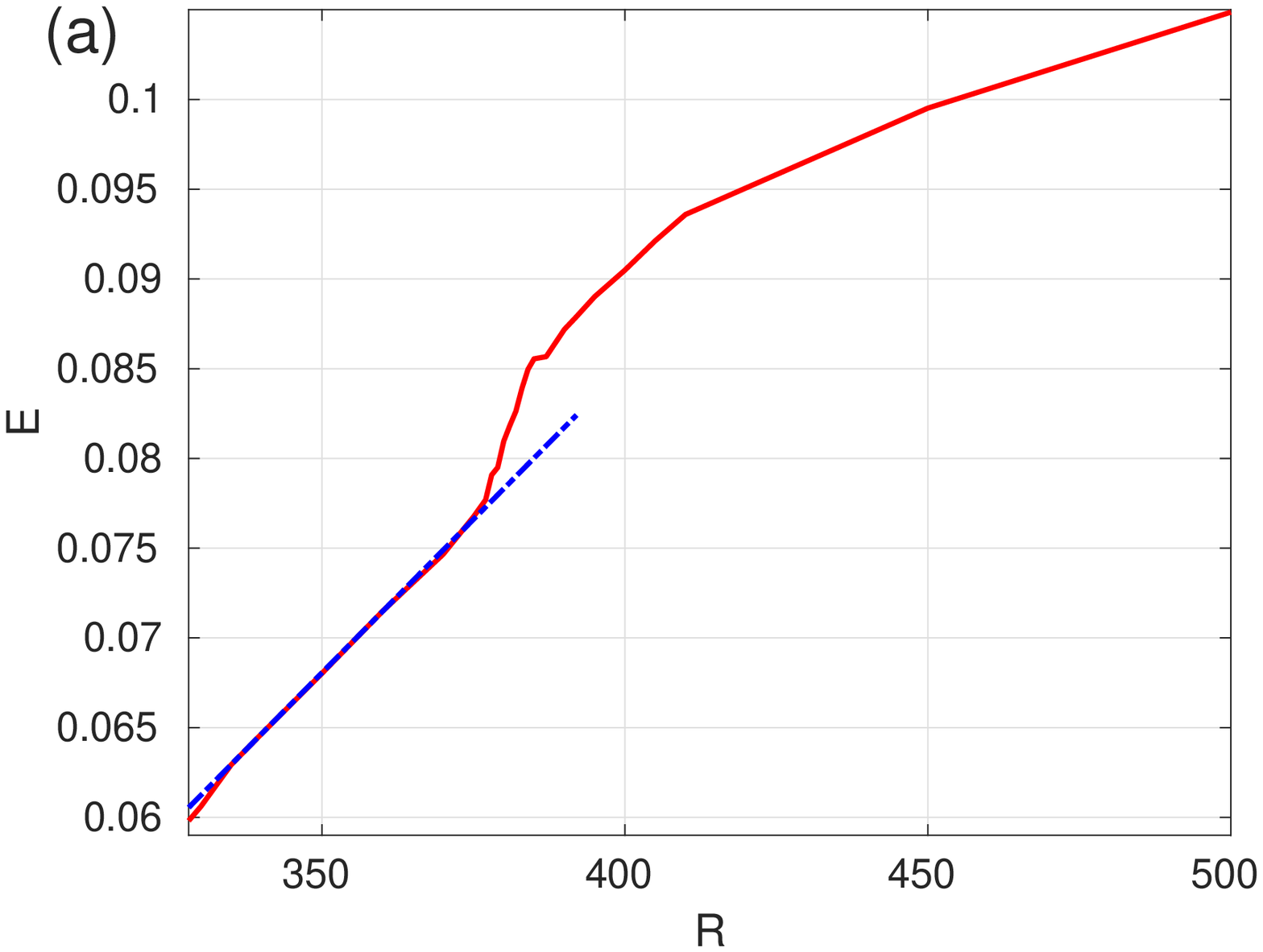}\includegraphics[width=6.5cm]{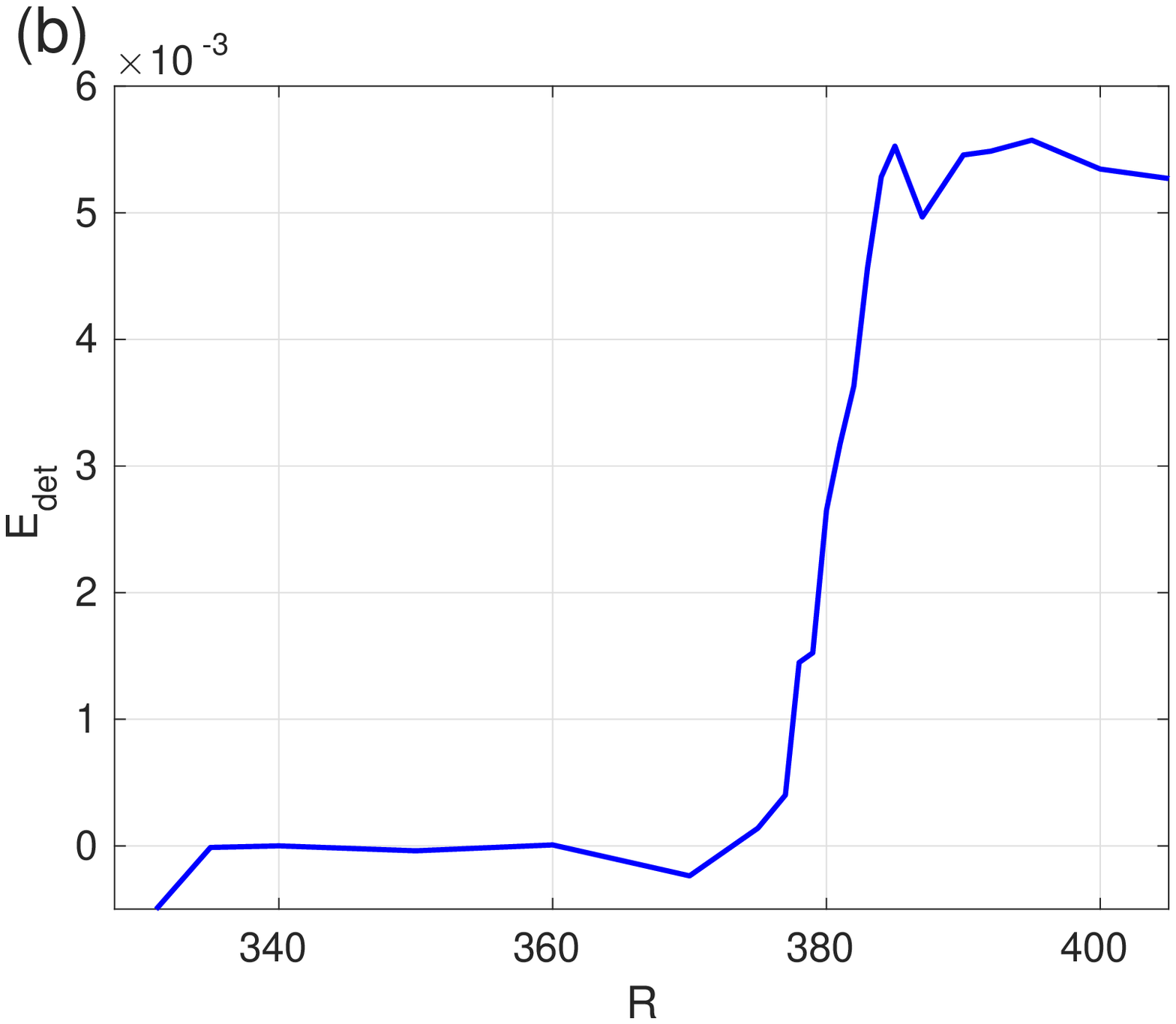}}
\centerline{\includegraphics[width=6.5cm]{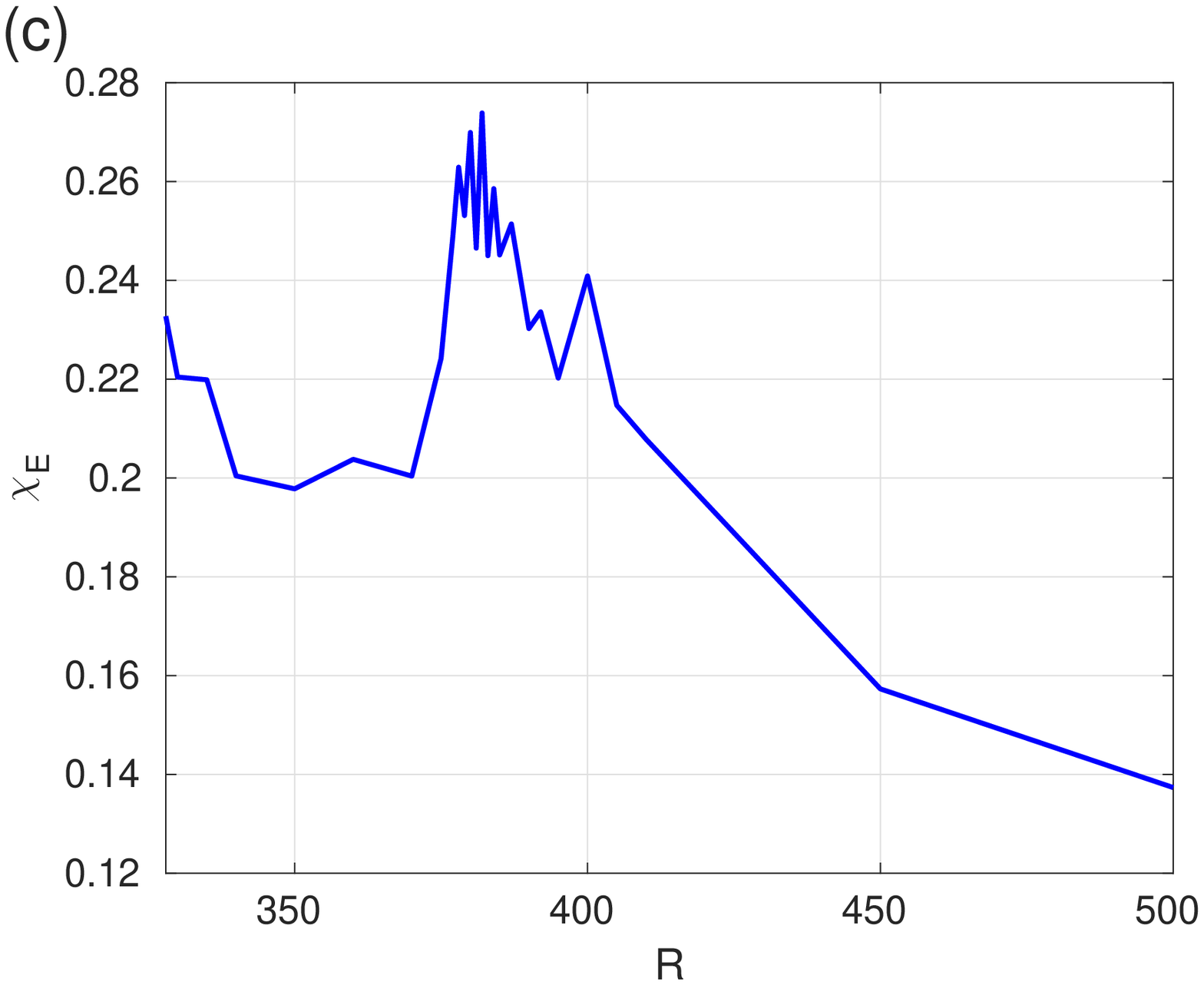}\includegraphics[width=6.5cm]{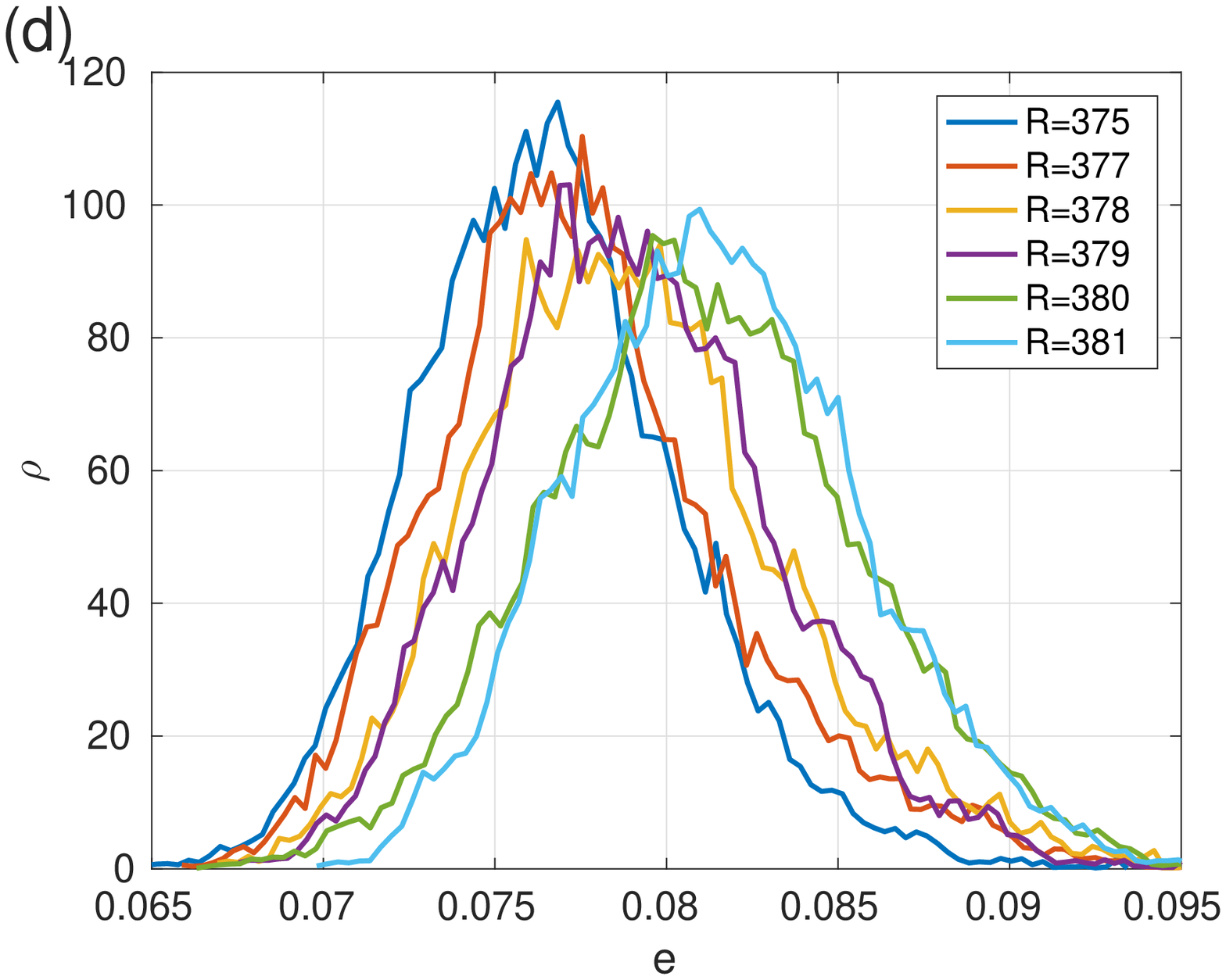}}
\caption{(a): Kinetic energy as a function  of the Reynolds number sampled in DNS in a domain of size $L_x\times L_z=110\times 32$.
The fit of $E(r), R<380$ by an affine function is superimposed as a dashed blue line. (b): Detrended kinetic energy $E_{\rm det}$ as a function of the Reynolds number sampled from the same simulations.
(c): Response function of the kinetic energy as a function of Reynolds number sampled from the same simulations.
(d) Probability density function of the kinetic energy for several Reynolds number in the range $[375;381]$.}
\label{figkin}
\end{figure}

\begin{figure}
\centerline{\includegraphics[width=6.5cm]{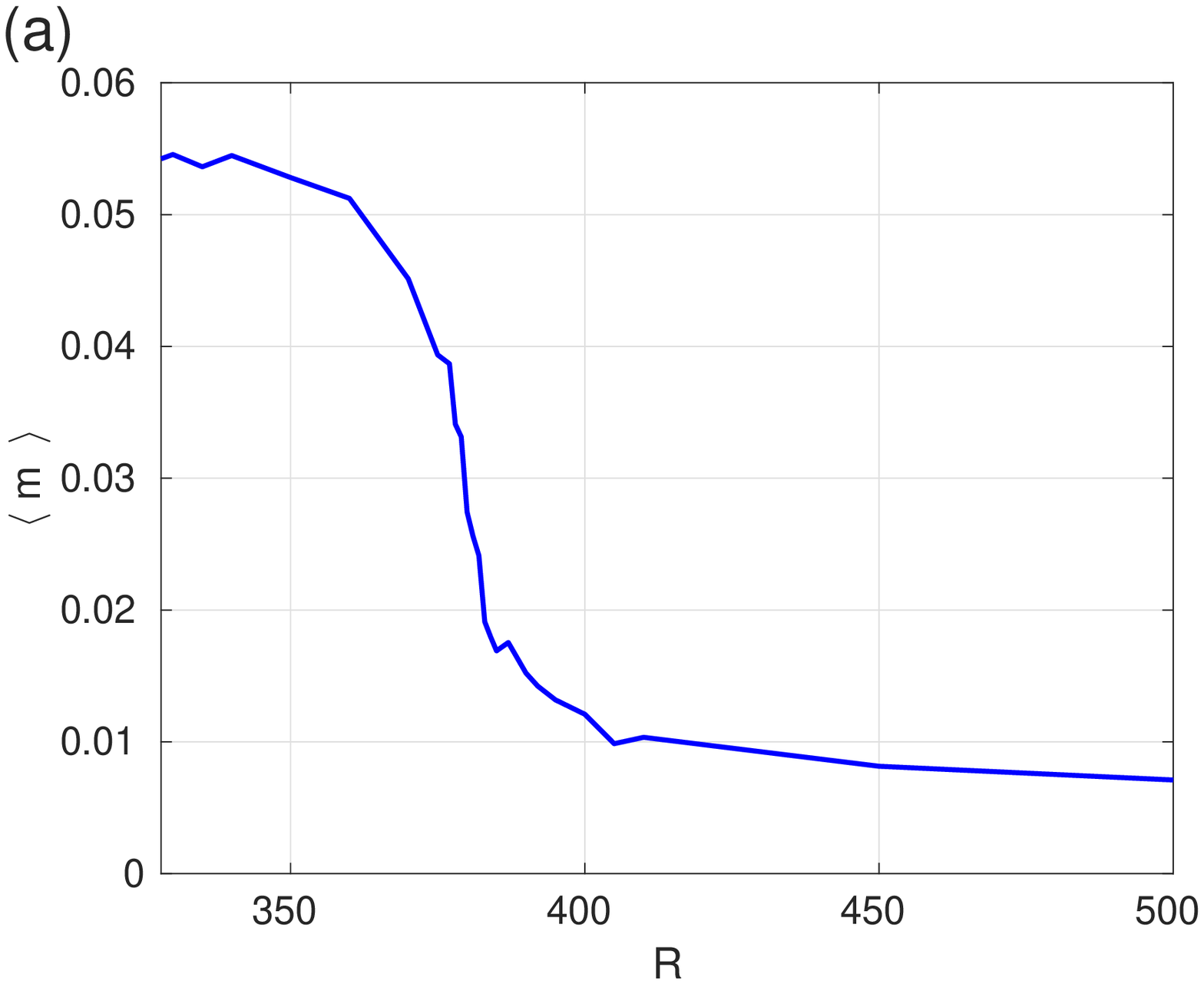}\includegraphics[width=6.5cm]{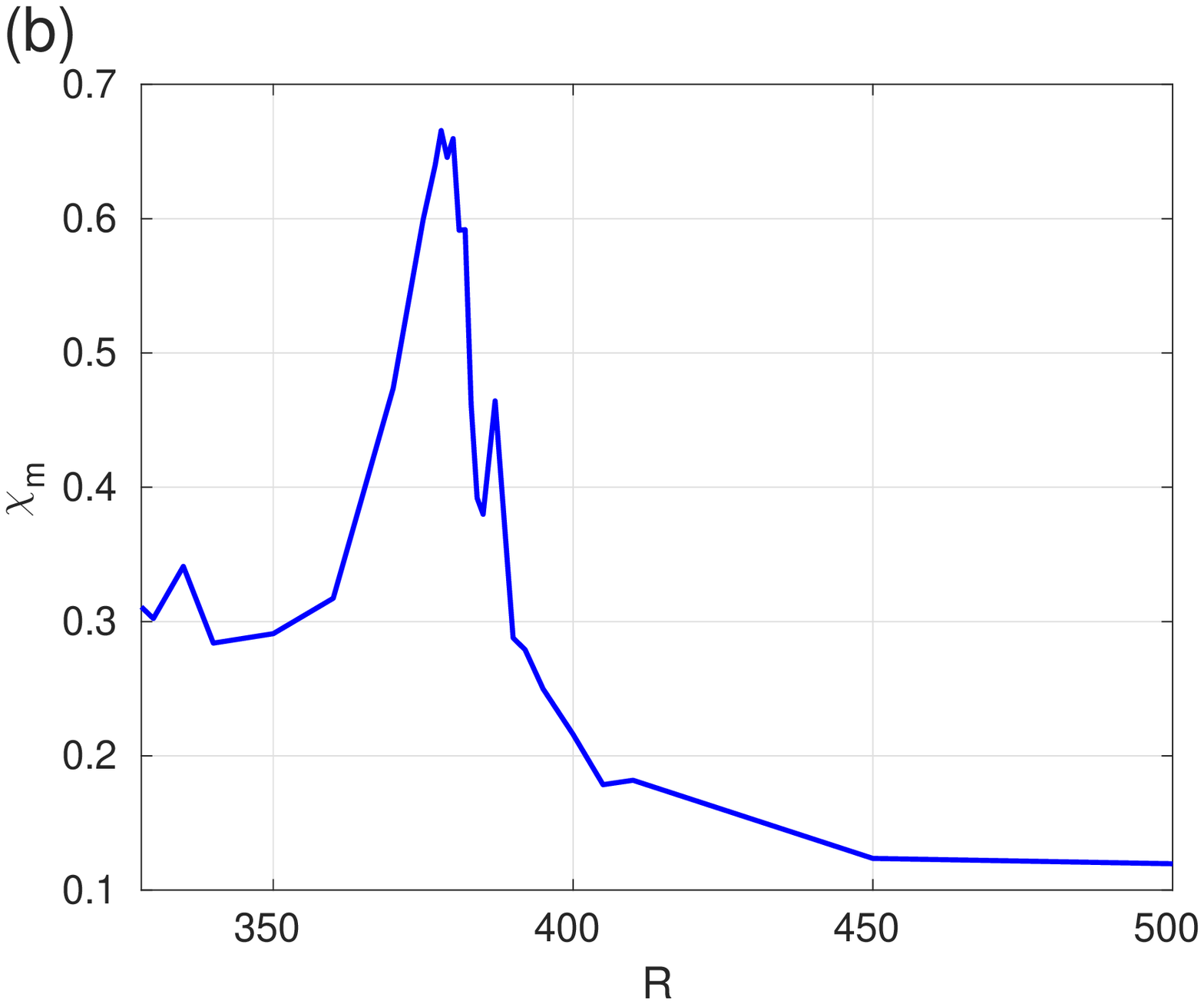}}
\caption{(a) : Average order parameter (amplitude of modulation of turbulence) as a function of the Reynolds number
sampled in DNS in a domain of size $L_x\times L_z=110\times 32$, containing one wavelength of the band.
(b) : Response of the order parameter in the same domain.}
\label{fig2}
\end{figure}

We present the kinetic energy and its response function sampled in DNS. Similarly to the order parameter,
the focus had so far been put on its dependence on the Reynolds number away from $R_{\rm t}$ \cite{RM10_1}.
The view on the range of Reynolds numbers close to $R_{\rm t}$ was even coarser for $E$ than it was for $M$. However,
visualisations of simulations both in small and large domain confirmed an earlier finding that uniformly turbulent
flow could coexist in time and in space with banded laminar turbulent coexistence \cite{RM10_1,BT11}. This had been
termed a reentrance of turbulence or intermittent regime. This is a further motivation to consider $E$ and $\chi_E$ closely in that range
of Reynolds number. We first comment on $E$ as a function of $R$ (Figure~\ref{figkin} (a)). We confirm the finding
that the kinetic energy increases like an affine function of $R$ for $R\lesssim 380$ (highlighted by the fit of $E$
with an affine function), and that it has a slower affine increase in $R$ for $R\gtrsim 400$ \cite{RM10_1}.
We now have a much finer sampling in Reynolds number in the range $R\in [380 ;390]$ and we can see that there is
a sharp increase between $R=378$ and $R=382$. In order to investigate this increase in more details,
we calculate the affine function which best fits $E$ for $R<375$: $E=aR+b$. Using these two coefficients,
we define a detrended kinetic energy $E_{\rm det}=E-(aR+b)$ for all Reynolds numbers (Figure~\ref{figkin} (b)).
For $E\le 375$, the detrended kinetic energy is nearly constant and close to zero, as expected from its definition.
For $R\ge 383$, again, the detrended kinetic energy  is nearly constant $E_{\rm det}\simeq 5\cdot 10^{-3}$.
Between $R=375$ and $R=383$ we can then detect  a jump of the detrended kinetic energy.
This definitely encourages us to use $E_{\rm det}$ to study a possible sharp crossover around $R_{\rm t}$
in the limit of infinite size. We then present the response function of the kinetic energy $\chi_E$ (Figure~\ref{figkin} (c)).
It shows the maxima in $380\le R \le 382$. The locus of these maxima is used to defined $R_{c, 2}$, at this resolution. This is not so surprising,
since the reentrance of turbulence is associated to kinetic energy moving between values corresponding to uniform
turbulence and banded laminar-turbulent coexistence, thus leading to larger fluctuations $\sigma_E$. We eventually present pdfs
of $e$ for Reynolds numbers in the range $R\in[375;381]$. The six curves can be grouped in three categories. At $R=375$ and $R=377$,
the pdfs are narrower and very similar, implying very close average, variance and response function of $E$: this corresponds to $E_{\rm det}\simeq 0$
and laminar-turbulent coexistence. At $R=380$ and $381$, again, the pdfs are narrower and very similar,
this corresponds to uniformly turbulent flow and $E_{\rm det}\simeq 5\cdot 10^{-3}$.
At $R=378$ and $R=379$, the pdfs are wider and lie in between their higher and lower Reynolds number counterparts. This corresponds to the maximum of the response function,
the jump of $E_{\rm det}$ and the intermittent reentrant turbulence regime.
While all these features indicate a clear qualitative change of state of the flow, quantified by the cumulants of kinetic energy,
this change of regime cannot be detected by examining the shape of probability density functions. While two states can be observed (bands and uniform turbulence),
all bimodality in the pdf of $E$ is an effect of undersampling and is erased by long enough time series. Missing such crossovers is a risk when studying a system solely
from the angle of $\mathbb{P}$-bifurcations. With the same dataset, we can display the order parameter $M$ (Fig.~\ref{fig2} (a)) and its response function (Fig.~\ref{fig2} (b)). This indeed shows the decrease of $M$ and the maximum of $M$ and the maximum of $\chi_M$ in the $R_{\rm t}$ region, seen in earlier studies. The locus of the maximum of $\chi_M$ can be used to define $R_{c,1}$ at this resolution. Both $R_c{c,1}\le R_{c,2}$ are contained in the looser $R_{\rm t}$ range where the change of regime between bands and uniform turbulence occurs. At this size, the features of $\chi_M$  and $\chi_E$ around their maxima are similar. This also shows us that we can finely determine the maximum of $\chi_M$.

\begin{figure}
\centerline{\begin{pspicture}(13,11)
\rput(4,8){\includegraphics[width=7cm,clip]{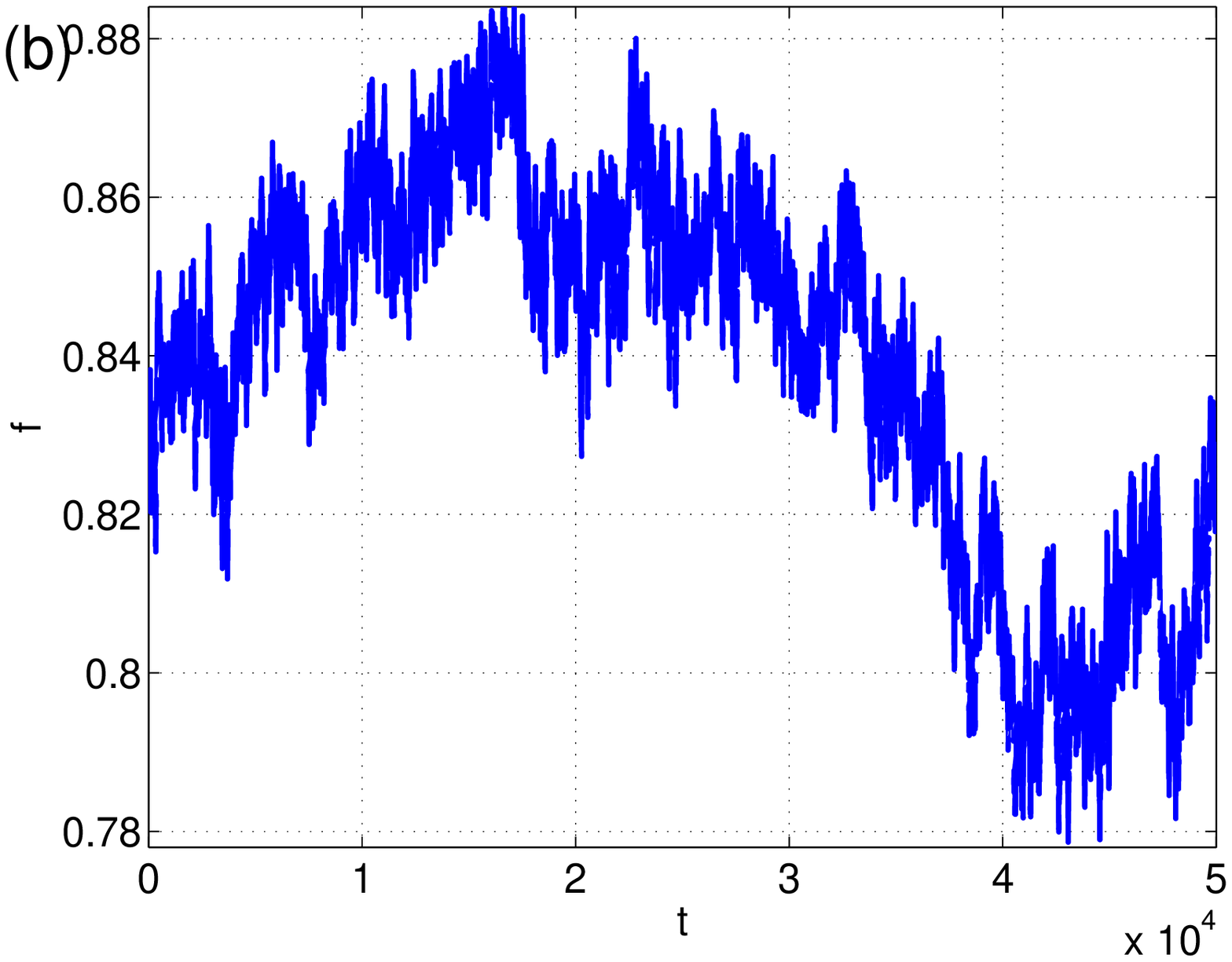}}
\rput(4,3){\includegraphics[width=7cm,clip]{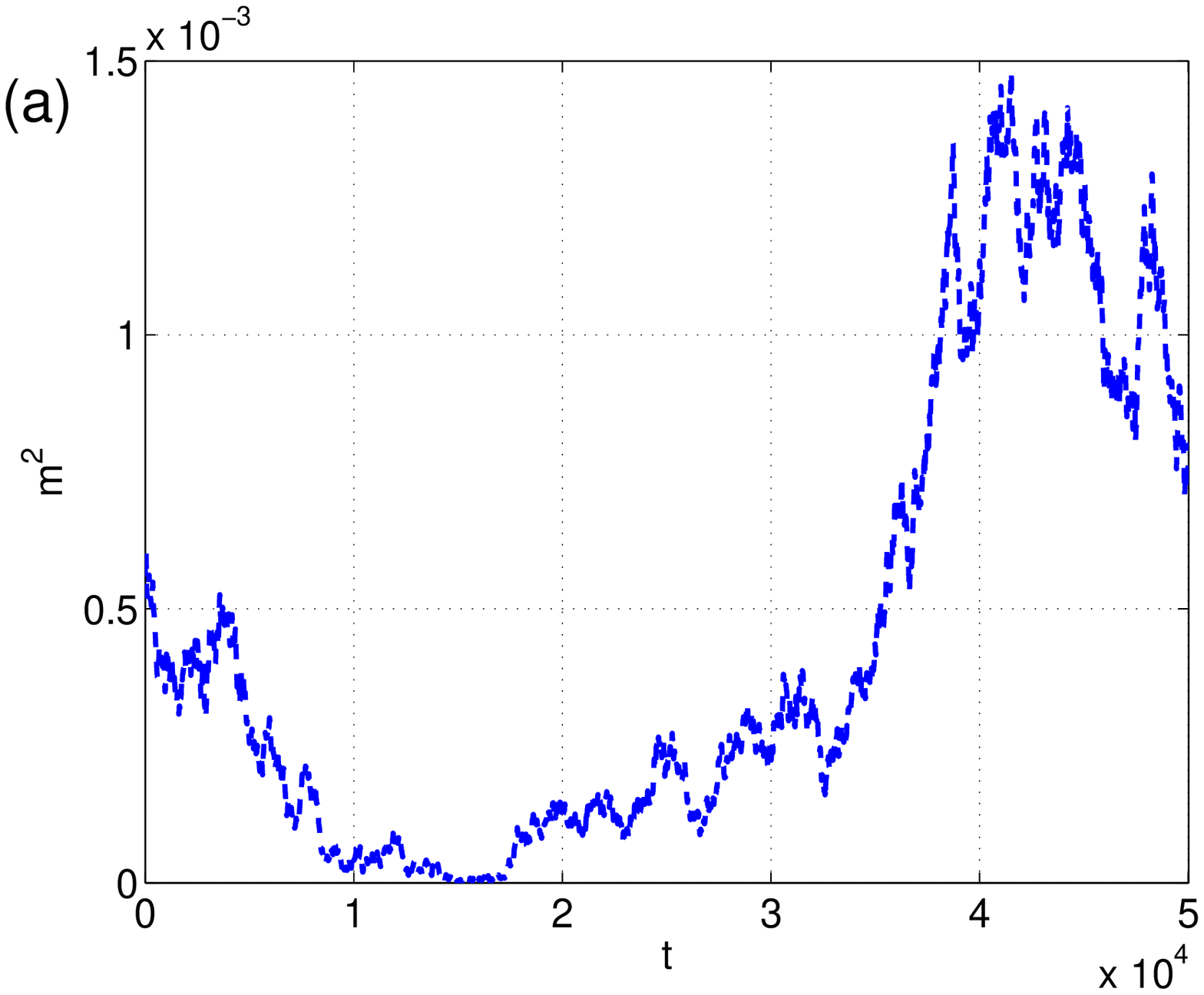}}
\rput(10,2){\includegraphics[width=5cm,clip]{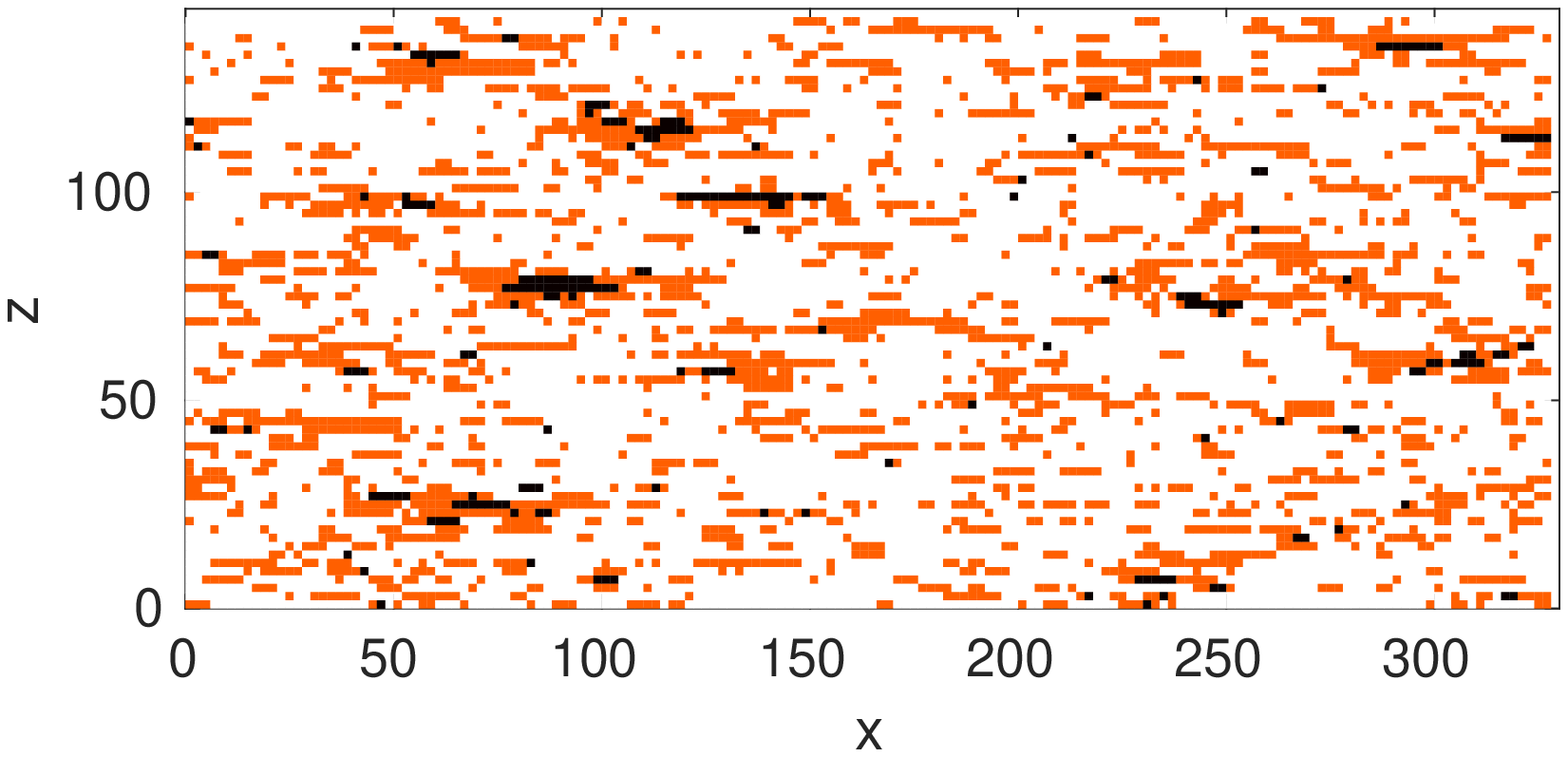}}
\rput(10,5){\includegraphics[width=5cm,clip]{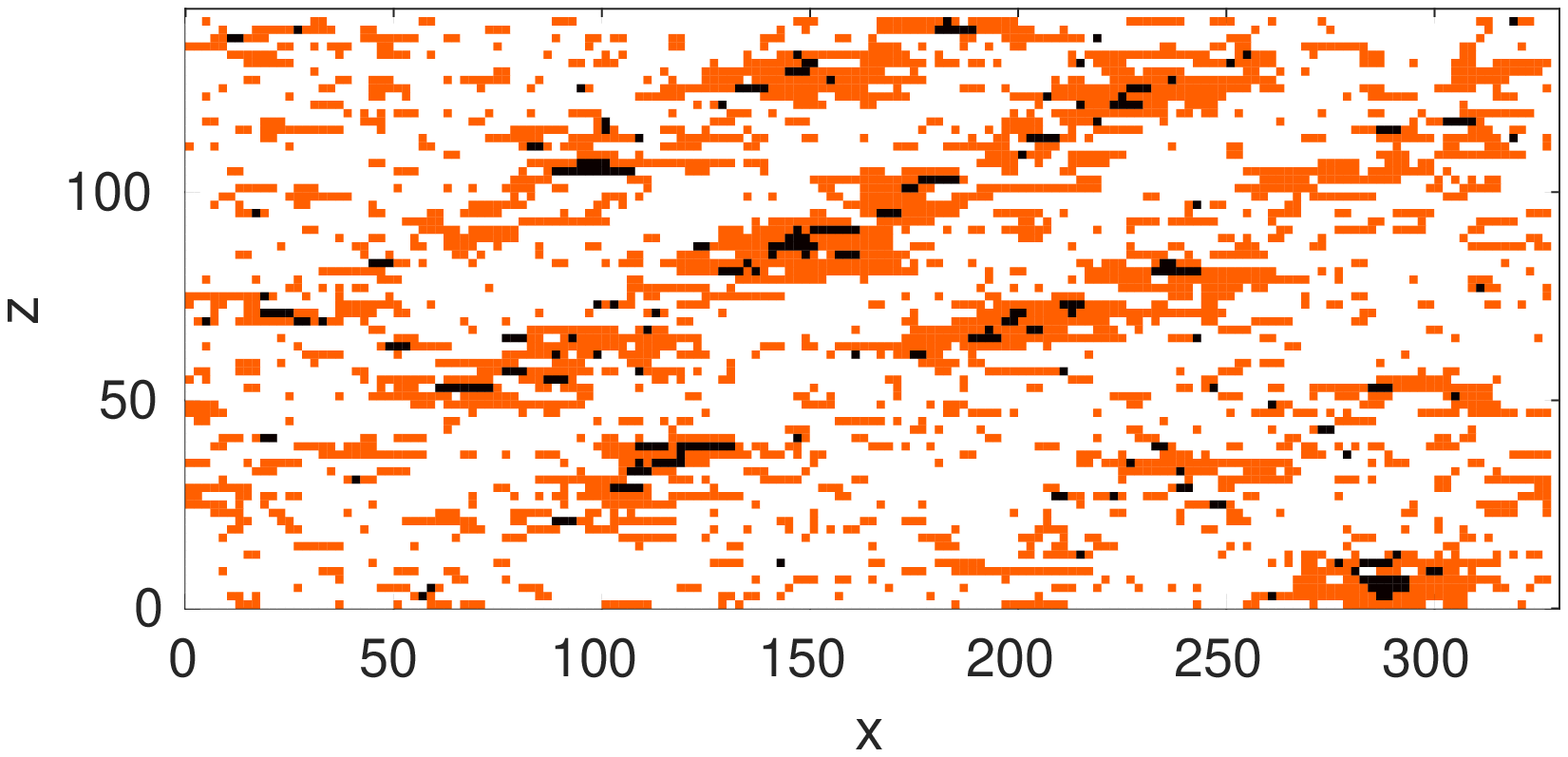}}
\psline[linecolor=blue](9.41,5.86)(8.15,5)
\psline[linecolor=blue](8.15,5)(9.4,4.2)
\psline[linecolor=blue](9.4,4.2)(12,5.3)
\rput(10,8){\includegraphics[width=5cm,clip]{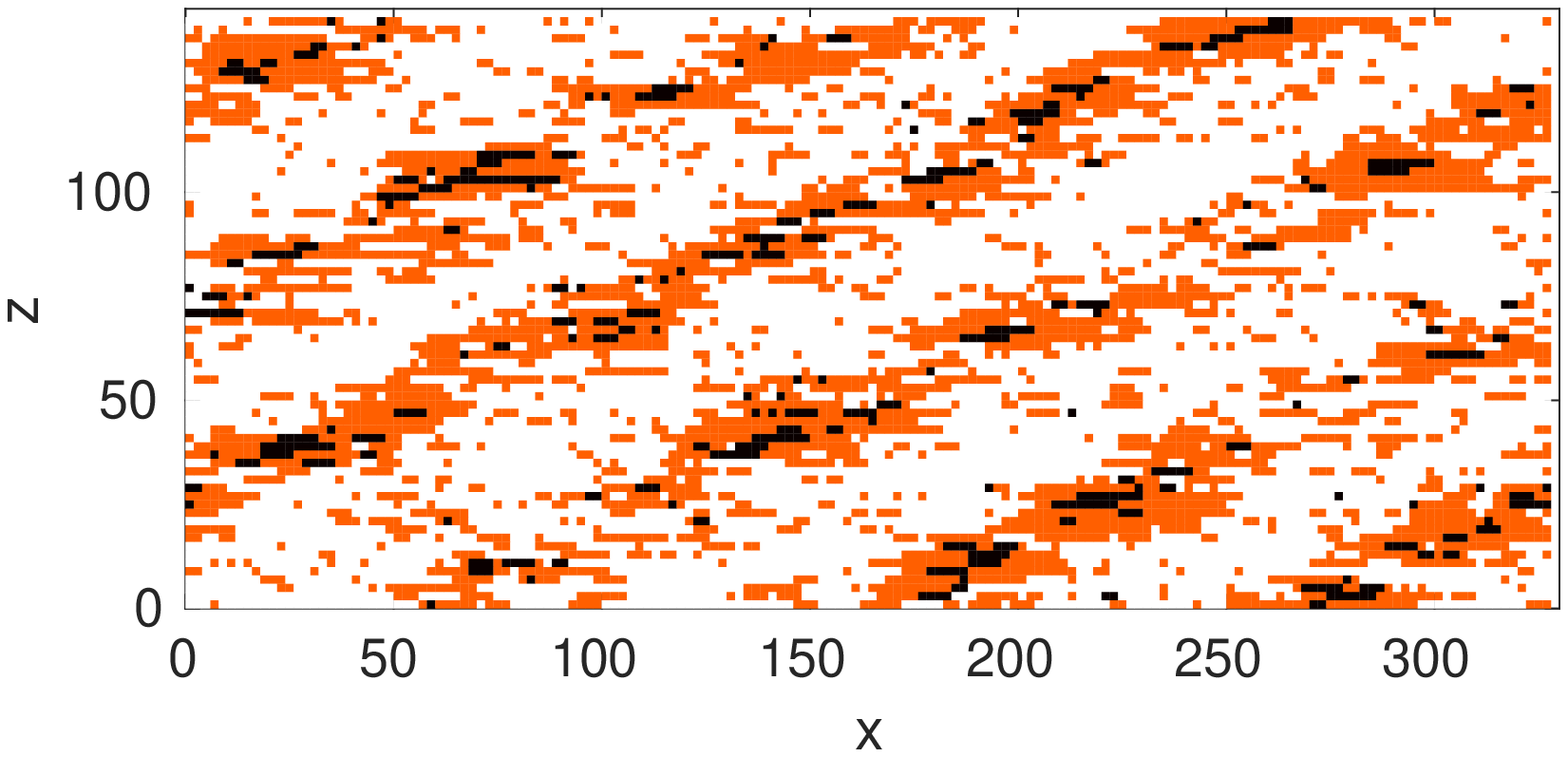}}
\psline[linestyle=dashed]{->}(3,1)(7.5,2)
\psline[linestyle=dashed]{->}(5.85,4.7)(7.5,8)
\psline[linestyle=dashed]{->}(5.16,2.42)(7.5,5)
\psline[linestyle=dashed]{<->}(3,1)(3,9.2)
\psline[linestyle=dashed]{<->}(5.85,4.7)(5.85,6.8)
\psline[linestyle=dashed]{<->}(5.16,2.42)(5.16,8.5)
\rput(10.3,5.2){\textcolor{blue}{$k_z=-4$}}
\rput(8.5,5.7){\textcolor{blue}{$T$}}
\end{pspicture}}
\caption{Illustration of reentrering turbulence in a domain of size $L_x\times L_z=330\times 144$ at $R=339.5$.
The figure comprises of time series of the order parameter (a) and the turbulence fraction (b), accompanied by
three typical snapshots of the Laminar/turbulent discriminated field at $t=15000$, $t=35000$ and $t=41000$.
The arrows link the instants in time to the corresponding snapshots.  The blue lines separate the domains occupies by
bands of different wavelength and orientation (indicated by the value and sign of wave vector $k_z$) or uniform turbulent indicated by $T$.}
\label{reen}
\end{figure}

Let us now consider a case of time series accompanied by visualisations of the flow in the form of the laminar-turbulent discriminated fields obtained in low order simulations.
This case concerns coexistence in time and in space of the band phase and the uniform turbulence phase.
Such a coexistence had already been seen \cite{RM10_1,BT11}. However, no thorough physical study of this regime had been performed.
We display the time series of turbulent fraction and order parameter illustrated by laminar-turbulent discriminated fields
in a system of size $L_x\times L_z=330\times 144$ at $R=339.5$ in figure~\ref{reen}. At the beginning of this example ($t\lesssim 10000$ )
laminar holes disappear from the flow, until it becomes uniformly turbulent ($10000\lesssim t\lesssim 25000$).
There are next to no laminar holes and no spatial organisation at large scale. This is indicated by the increasing then
plateauing turbulent fraction, and the very small values reached by the order parameter. This can be illustrated by a
laminar-turbulent discriminated field at $t=15000$. After some time, the turbulent fraction decreases and the order parameter increases:
laminar-turbulent bands reappear locally in the flow. This is illustrated by a laminar-turbulent discriminated field at $t=35000$.
Fronts between the area where laminar-turbulent bands are found and the area where uniform turbulence is found slowly changes.
The banded state eventually invades the whole flow: this is illustrated at $t=41000$ by the laminar-turbulent discriminated field.
Note also that the turbulent fraction has reached a low plateau and the order parameter a higher plateau.

\section{The finite size analysis}\label{res}

\subsection{The double transition}

\begin{figure}
\centerline{\includegraphics[width=7cm]{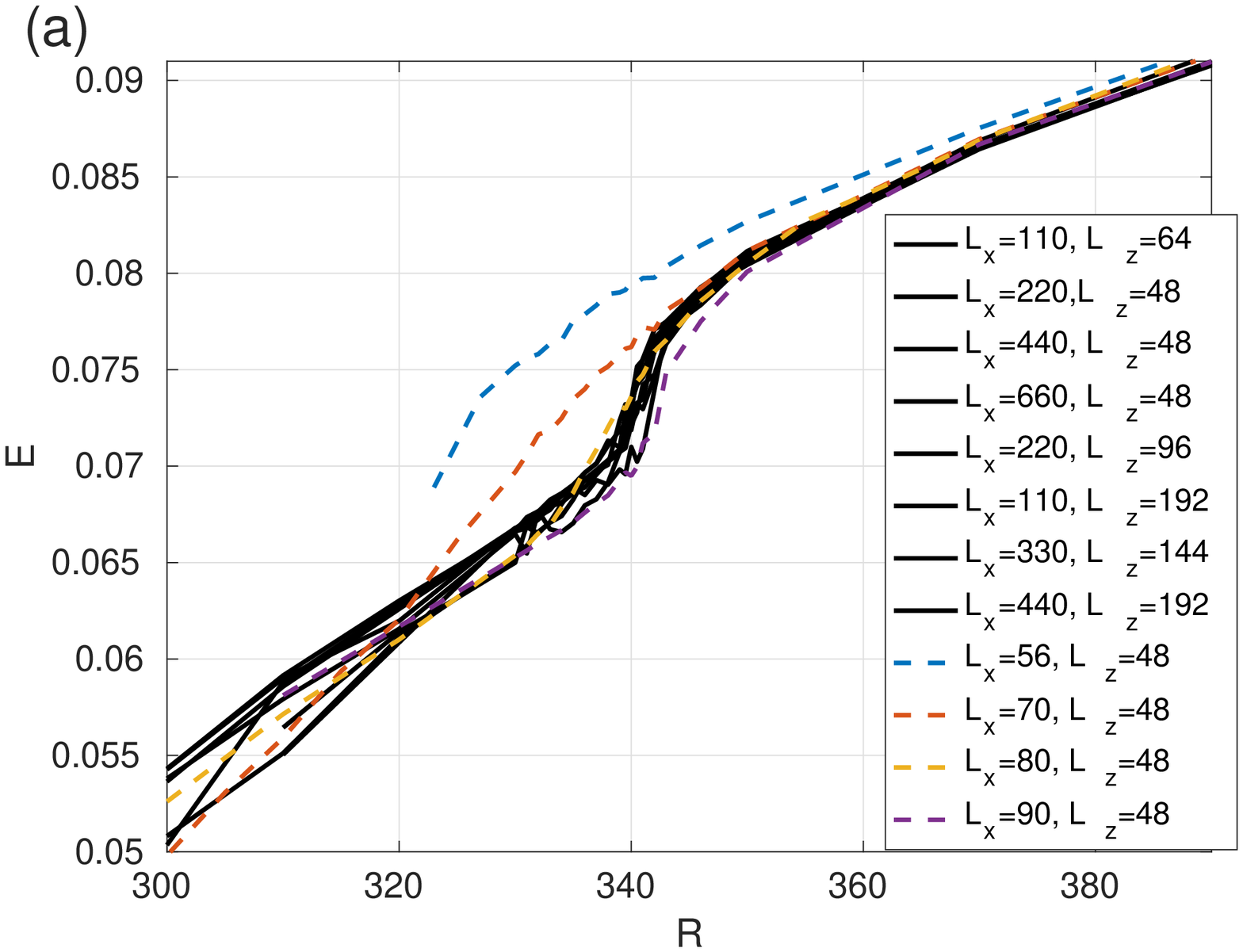}\includegraphics[width=7cm]{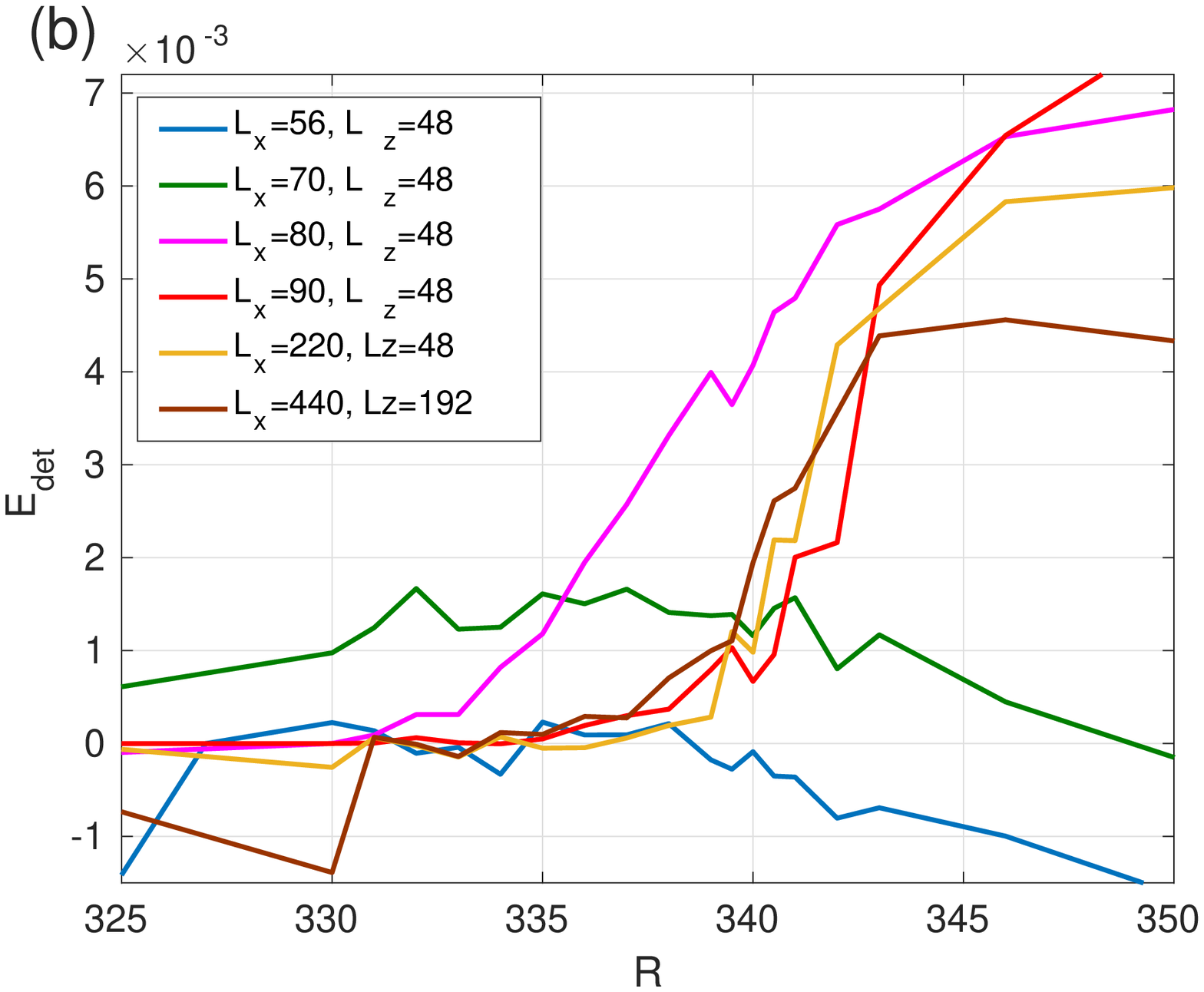}}
\centerline{\includegraphics[width=7cm]{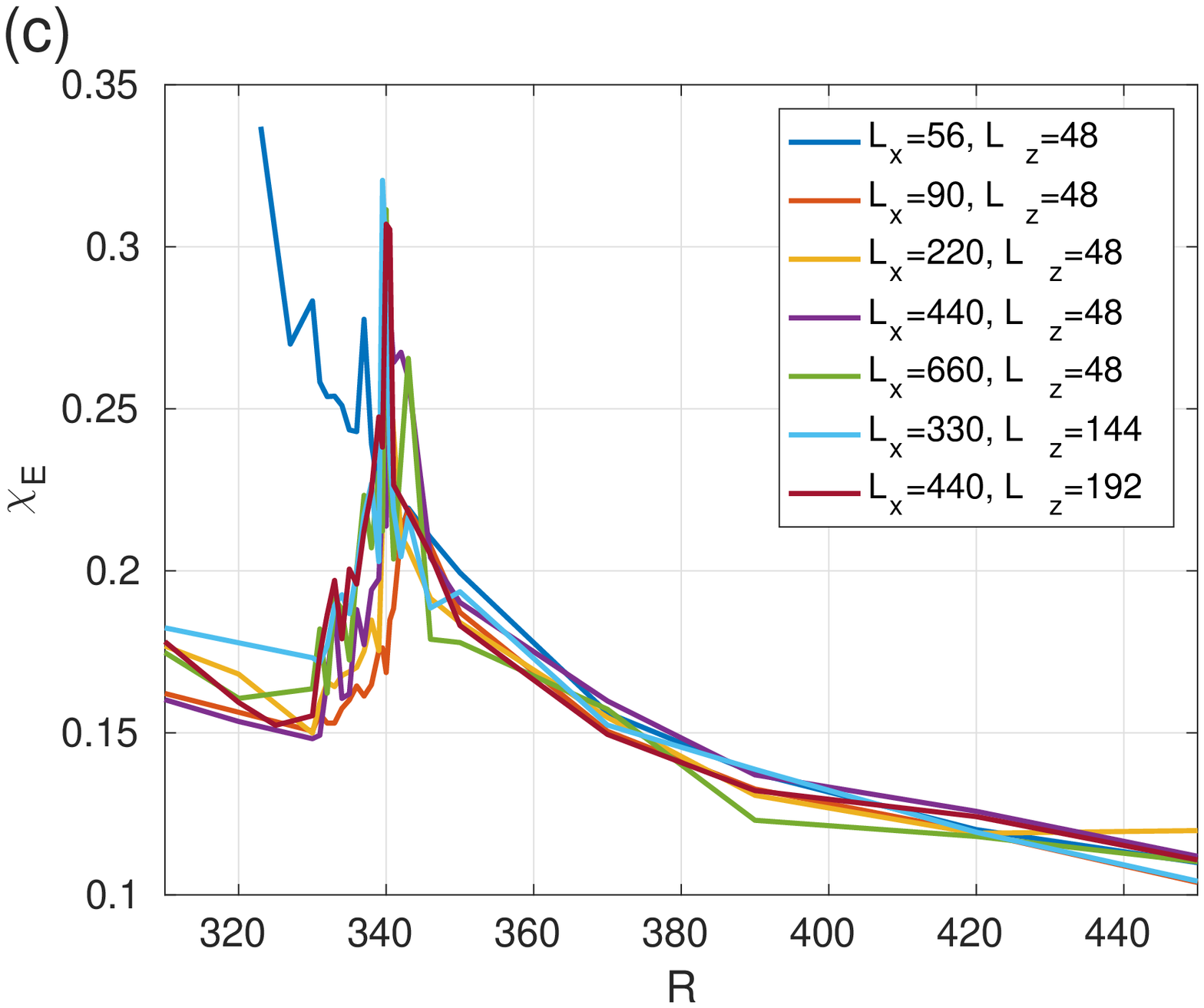}}
\caption{(a): Average kinetic energy as a function of the Reynolds number for all the considered domain sizes.
(b): Detrended average kinetic energy as a function of Reynolds number for the smaller domain sizes, zoomed on the transition range.
(c): Response function of the kinetic energy as a function of the Reynolds number for all the considered domain sizes.}
\label{KE}
\end{figure}

  We first examine the dependence of $E$ on the Reynolds number for increasing domain sizes. We sampled $E$ in low order
  simulations over a large range of Reynolds numbers which always include $[310 ; 450]$ for domain sizes going from
  $L_x\times L_z= 56\times 48$ to $L_x\times L_z=440\times 192$. The results are displayed as a function of the
  Reynolds number in figure~\ref{KE} (a). For domains of size larger than or equal to $L_x\times L_z=110\times 64$ all the values
  (in black) collapse on a master curve. This curve shows three ranges of Reynolds number. For $R\le 338$,
  $E$ increases like an affine function of $R$, as was seen in earlier simulations \cite{RM10_1} and in
  DNS (Figure~\ref{figkin} (a)). for $R\ge 340$, we find once again the slower increase of $E$ with $R$ already
  seen in the uniformly turbulent phase. In the narrow range $338\le R\le 340$, one can see a very rapid increase of $E$,
  which joins the two regimes. Finite size analysis shows the convergence of $E(R)$ as size is increased
  from $L_x\times L_z=56\times 48$ to $L_x\times L_z=110\times 64$: the curves are closer and closer to the asymptotic, large size, state if $R\le 350$.
    Meanwhile, one finds very little difference in $E(R)$ in the uniformly turbulent phase if $R\ge 360$.

We then focus on the narrow though important range of Reynolds number $338\le R\le 340$. For this,
we will extensively use the detrended kinetic energy $E_{\rm det}$. The calculation is performed independently
for each domain and it is similar to what is performed in section~\ref{phe}. We first fit $E(R)$ by an affine
function $aR+b$ in the relevant range of $R\in [310; 337]$. In the system of size $L_x\times L_z=56\times 48$, we use the different range
 $R\in[327;337]$. We then have $E_{\rm det}(R)=E(R)-(aR+b)$. We display $E_{\rm det}(R)$
for increasing domain sizes in figure~\ref{KE} (b). For domains of size larger than $L_x\times L_z=110\times 64$
(termed ``large size'' domains), we recover the situation seen in direct numerical simulations: $E_{\rm det}$ remains
nearly zero up to $R\lesssim 338$. For $R\gtrsim 342$, the detrended kinetic energy is again constant at $E_{\rm det}\simeq 5\cdot 10^{-3}$.
The range of Reynolds numbers $[338;340]$ is where the reentrance of turbulence occurs. The spatial and temporal coexistence of uniformly
turbulent flow and laminar-turbulent bands occurs. Note that bimodality is again absent from the pdfs of $e$.

The situation is quite different for the smaller domains. In the domain
of size $L_x\times L_z=56\times 48$, $E_{\rm det}$ is constant for $R\le 327$. As $R$ is decreased to smaller values,
$E$ and $E_{\rm det}$ decrease very fast: the domain can completely relaminarise. This domain size is very peculiar,
no organised laminar-turbulent coexistence can be found: below $R=345$, one only has very intermittent formation,
shape shifting and disappearance of laminar holes. A hint of the ``large size'' behaviour can be seen in the lager
domain $L_x\times L_z=70\times 48$: $E_{\rm det}$ is nearly zero up to $R=330$ and then jumps to $E_{\rm det}\simeq 10^{-3}$
for higher values of $R$. We then consider a larger domain of size $L_x\times L_z=80\times 48$: in that case,
$E_{\rm det}$ is near zero up to $R\simeq 334$, increases slowly up to $6\cdot 10^{-3}$ and remains on this
plateau for $R\gtrsim 343$. In the last ``small size'' domain considered ($L_x\times L_z=90\times48$), $E_{\rm det}$
is nearly indistinguishable from what is sampled in the ``large size'' domains. From this, one can draw a size
dependent picture of the increase of $E_{\rm det}$ with $R$. As soon as a somewhat steady laminar-turbulent
coexistence is possible ($L_x\times L_z\ge70\times48$), $E_{\rm det}$ grows from a plateau value to another as $R$ is increased.
As the size of the domain is increased, the steepness of the jump increases and the Reynolds number at which it occurs converges toward $R\simeq 340$.
Note that the establishment of the crossover between laminar-turbulent coexistence and uniform turbulence as domain size is increased had
been seen in earlier simulations, with less emphasis on quantitative measurements \cite{pm11}.

We then present the response function $\chi_E$ as a function of Reynolds number for some of the domains in
which we performed simulations (Figure~\ref{KE} (c)). Two types of behaviour are visible. For the smallest system
$L_x\times L_z=56\times 48$, the response function increases monotonously as $R$ is decreased and shows no sign of
a crossover until relaminarisation become very probable around $R=323$. This is typically the trace of the very intermittent
laminar-turbulent coexistence without spatial organisation, as already seen in small size domains \cite{pm11}. As soon as the system
is large enough for a spacial organisation in bands to be possible (\emph{i.e.} provided the domain is larger than $L_x\times L_z=90\times 48$), the response function reaches an asymptotic behaviour.
The function $\chi_E$ first increases as $R$ is decreased and has a clear maximum in the narrow range $R=340\pm3$.
A large majority of maxima of $\chi_E$ is actually in the range $R=340\pm 1$. Note that this maximum corresponds to a spike of $\chi_E$ and that it is reached precisely at the Reynolds number
where one saw the steep augmentation of $E_{\rm det}$ (Figure~\ref{KE} (b)). Another peculiar fact is that the value
of the maximum of $\chi_E$ depends very little on size: all the values sampled fall within $20\%$ of the mean. There
is also no clear monotonous dependence on the size $L_x$, $L_z$ or $L_xL_z$. We can thus define $R_{c,2}=340\pm 1$ as $\lim_{L_xL_z\rightarrow \infty}\arg\max_{R}\chi_E$, which belongs to the range $R_{\rm t}\simeq 345\pm 10$ at this resolution.  As $R$ is further decreased, $\chi_E$ decreases smoothly.

\begin{figure}
\centerline{\includegraphics[width=7cm]{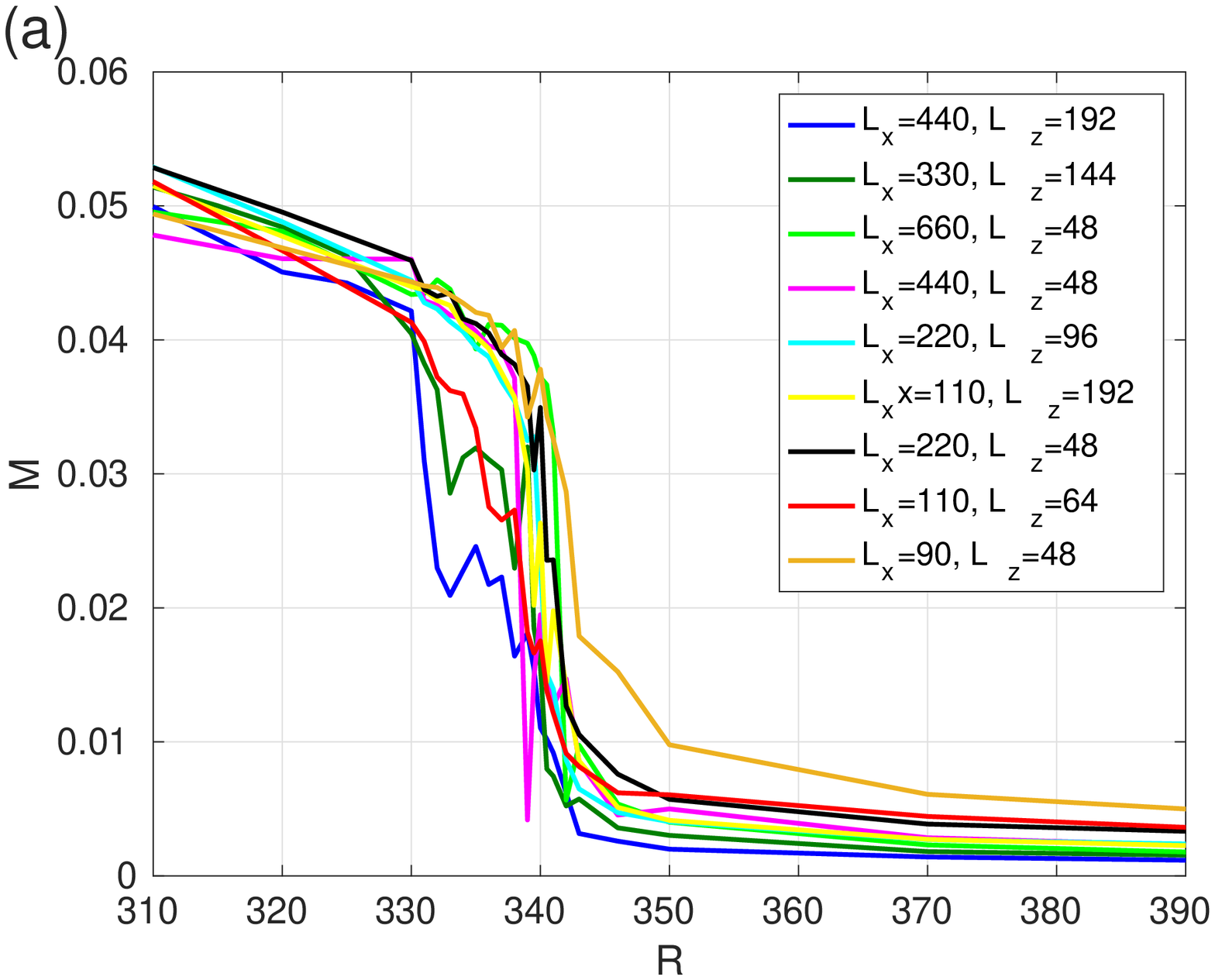}\includegraphics[width=7cm]{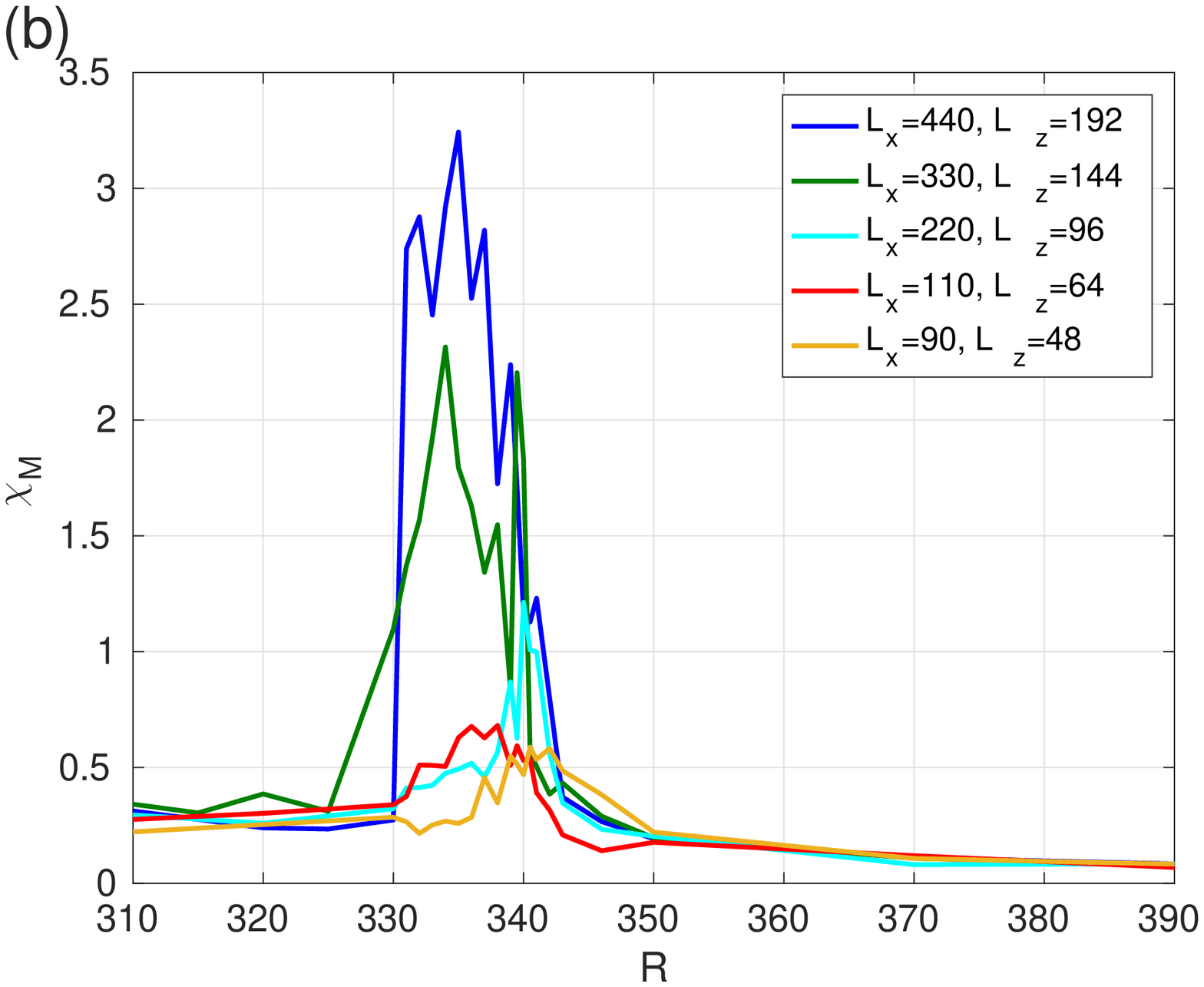}}
\caption{(a): Order parameter as a function of the Reynolds number for all the considered domain sizes.
(b): Response function of the order parameter as a function of the Reynolds number for the considered domain sizes.}
\label{mchi}
\end{figure}

We then present the Reynolds number and domain size dependence of the order parameter $M$ and its response function $\chi_M$.
The order parameter is displayed as a function of $R$ in figure~\ref{mchi} (a) for domains of size ranging from $L_x\times L_z=90\times 48$
to $440\times 192$. The data sampled bridges the gap between the view of Taylor Couette flow in a very large domain considered
by Prigent \emph{et al.} \cite{prigent02,pr_physD} and the earlier simulations of plane Couette flow in a periodical domain containing
one wavelength of the bands \cite{RM10_1} (see also \S~\ref{phe}). For $R\le 330$, $M(R)$ depends very little on size:
we find a concave decrease of $M$ in agreement with the square root decrease found in earlier experiments and numerical simulations.
The order parameter further decreases in the range $330<R<340$. There is no clear dependence yet on the domain size for $M(R)$,
even if the data sampled in the domain $L_x\times L_z=440\times 192$ lies out. This is a consequence of the large
 relative fluctuations of $m$. For $R>340$, $M(R)$ decreases smoothly with $R$ in all domains. For these Reynolds numbers,
 the modulation of turbulence is not visible any more. A precise zoom on this range actually reveals a decrease of $M$ with the domain size. This assertion and the precise scaling in size and Reynolds number will be checked in section~\ref{dis}. This is consistent with the near zero values found experimentally in a very large Taylor--Couette apparatus \cite{prigent02}.

We now consider the response function $\chi_M$ as a function of the Reynolds number. The processed data is presented in figure~\ref{mchi} (b)
for domain sizes ranging from $L_x\times L_z=90\times 48$ to $L_x\times L_z=440\times 192$. For the smaller Reynolds numbers $R\le 330$,
$\chi_M$ is independent of size and depends very little on the Reynolds number. For the larger Reynolds numbers $R\ge 346$,
the response function is again independent of size and decreases smoothly with the Reynolds number.
The scaling in Reynolds number of this decrease will be considered in more details in section~\ref{dis}.
For this quantity, the range of Reynolds number of interest is $R\in[330; 346]$. The data sampled here goes beyond
the view given by one domain size (see \cite{RM10_1} and \S~\ref{phe}). We can see that $\chi_M$ really distinguish itself
from $\chi_E$ (Figure~\ref{KE} (c)) in its dependence on domain size. Indeed, $\chi_M$ takes large values over the whole range
 $R\in [330;340]$, not just in a narrow spike. Each curve, sampled in a domain of a given size can be distinguished from all the others. Stated more precisely,
 $\chi_M^m(L_x,L_z)\equiv \max_R\chi_M$ grows monotonously with the size without any ambiguity. Because of the complexity of the type
 of fluctuations (orientation fluctuations, wavenumber selection, reentrance of turbulence) and the relatively large variance of $M$, extremely large time series would be required to replace the seesaw maxima of $\chi_M(R)$ in the range $R\in [330;340]$ by clearly distinguishable peaks.
 In particular, in data sampled in the domain of size $L_x\times L_z=330\times 144$, visualisations and time series of $m$ and $f$ of the type of figure~\ref{reen} indicate that the large values of $\chi_M$ near $R= 334$ correspond to orientation fluctuations and the large values of $\chi_M$ near $R= 339$, are greatly influenced by reentering turbulence
 on top of orientation fluctuations. These two peaks are also visible in $\chi_M$ sampled in the domain of size $L_x\times L_z=220\times 96$.
 However, for most domains of smaller than $330\times 144$, the effect of orientation fluctuations and reentering turbulence are mixed
 in the maxima of $\chi_M$. The situation is more complex in the largest domain $L_x\times L_z=440\times 192$: while $\chi_M$ is larger than in other domains, it has rapid changes with Reynolds numbers.

\begin{figure}
\centerline{\includegraphics[width=7cm]{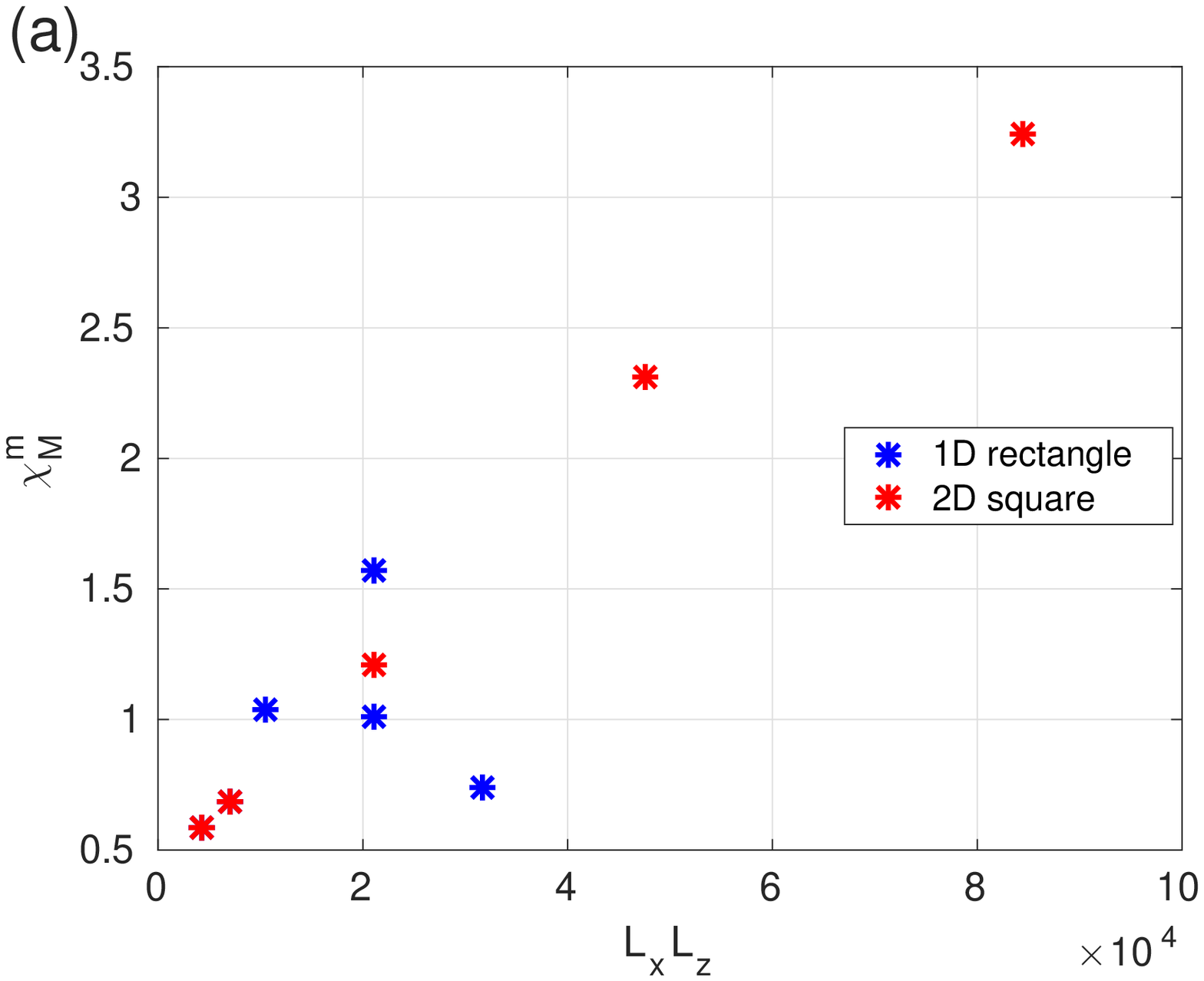}\includegraphics[width=7cm]{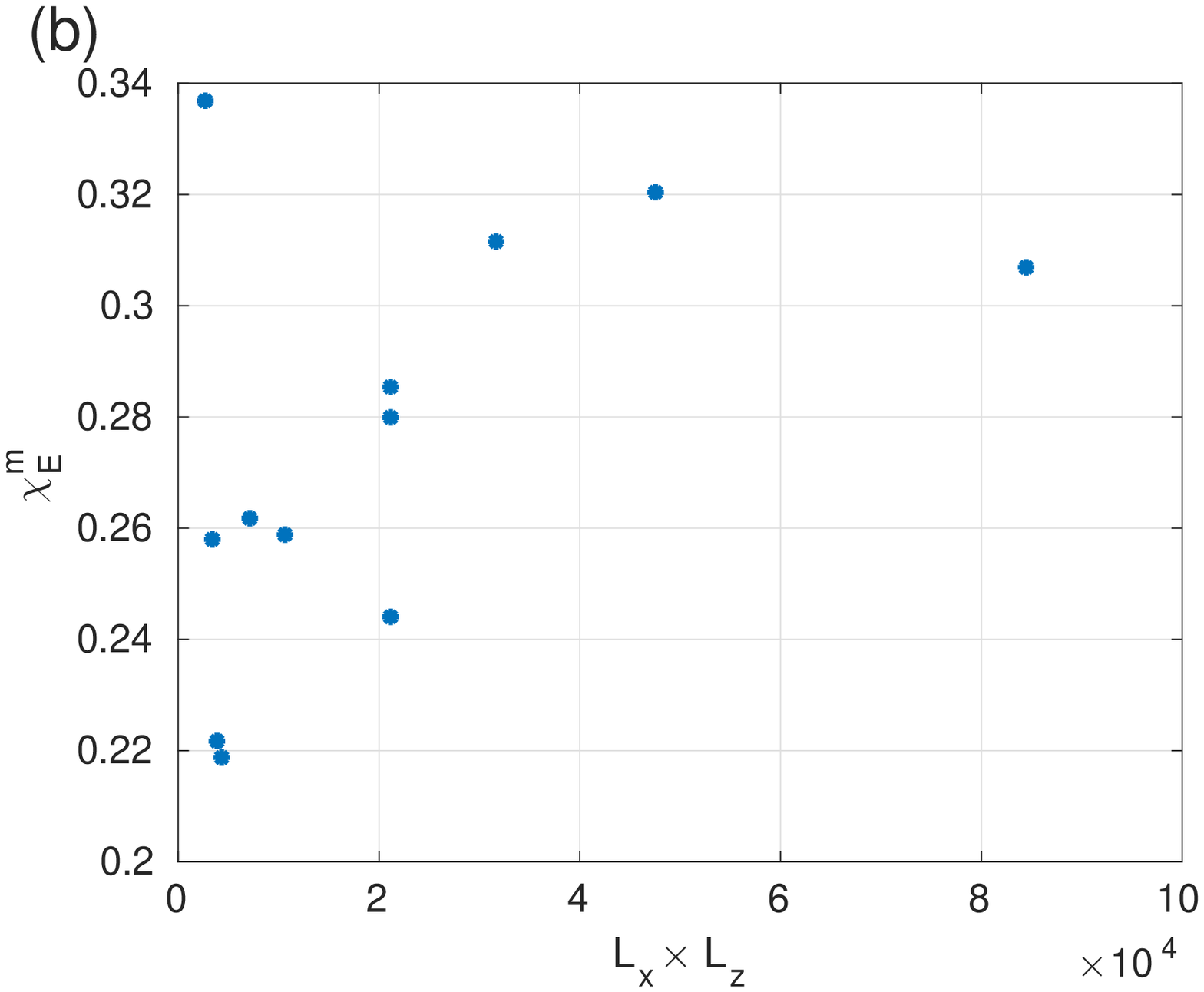}}
\caption{(a): Maximum over Reynolds number of the response function of the order parameter as a function of system surface.
 The colour distinguishes the square domains where $n_x/n_z=1$ and the rectangle domains where $n_x/n_z\ne 1$
(b): Maximum over Reynolds number of the response function of kinetic energy as a function of system surface.}
\label{fsa}
\end{figure}

In order to follow more clearly the increase of $\chi_M(R)$ with the size, we focus on $\chi_M^m(L_xL_z)$ as a function of $L_x\times L_z$, displayed in figure~\ref{fsa} (a).
the Reynolds number at which the maximum is reached,  $R_L\equiv \arg \max_M \chi_M$, is also calculated (not shown here). This follows the spirit of finite size analysis (\S~\ref{intro}, Eq.~(\ref{FSA_i})).
The growth of $\chi_M^m$ is clearly visible. The monotonic, regular, growth in the range of size $7\cdot 10^4\le L_xL_z\le 8.4\cdot 10^5$ (which covers more than a decade) is entirely compatible with a the power law behaviour
type. The data acquired here cannot yield
an extremely precise value as to the ratio of exponents $\tilde{\nu}/\mu$. However, it does indicate that the order parameter $M$ undergoes a second order phase transition, even more clearly than an estimate in a domain of a single size, no matter how large. Moreover, the estimate given here is much more precise than a direct calculation from a fit of the response function (Figure~\ref{mchi} (b)) Finite size analysis thus improves greatly the quality of numerical studies of phase transition at a given computational cost.
 Note that $\chi_M^m$ grows with size, but depends little on the aspect ratio $L_x/L_z$,
as can be seen in the three domains of size $L_x\times L_z=110\times 192$, $220\times 96$ and $440\times 48$. We considered $R_L(L_xL_z)$.
The data acquired shows that this Reynolds number is included in the range $[334;341]$ and is larger for small domains than it is for large domains. Much longer time series would be required to confirm precisely the tendency of variation of $R_L$. Provided that we could add larger datasets in a handful of larger domains, we could precisely define $R_{c,1}=\lim_{L_xL_z}R_L$. We systematically find that $R_L\lesssim R_{c,2}$, so that we expect that $R_{c,1}\le R_{c,2}$. All three Reynolds numbers $R_L$, $R_{c,1}$ and $R_{c,2}$ belong to the $R_{\rm t}$ range.
We eventually present the size dependence of the maximum of the response function of the kinetic energy $\chi_E^m$ in figure~\ref{fsa} (c). This
shows how the asymptotic regime is reached by the cumulant of $E$. More importantly, this shows that the crossover type underwent by kinetic energy
is not a first order transition. Such a transition would mean that $\chi_E^m$ grows indefinitely and that $\chi_E$ is narrower and narrower as size is increased. Again, this is another improvement brought by the finite size analysis, since a study in a domain of a single size could have mislead one into thinking that the crossover was a first order transition.

\subsection{The disordered phase}\label{dis}

\begin{figure}
\centerline{\includegraphics[width=7cm]{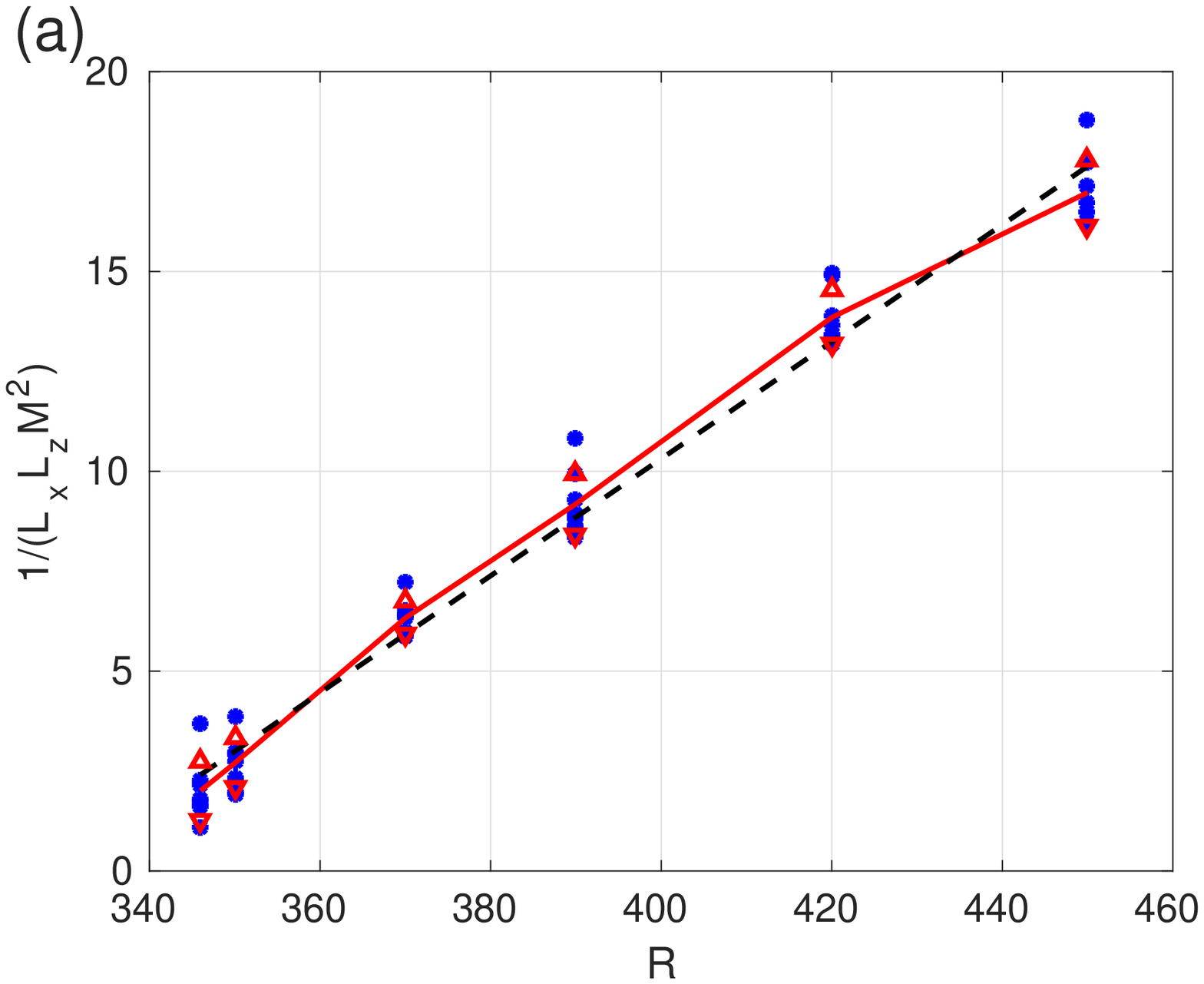}\includegraphics[width=7cm]{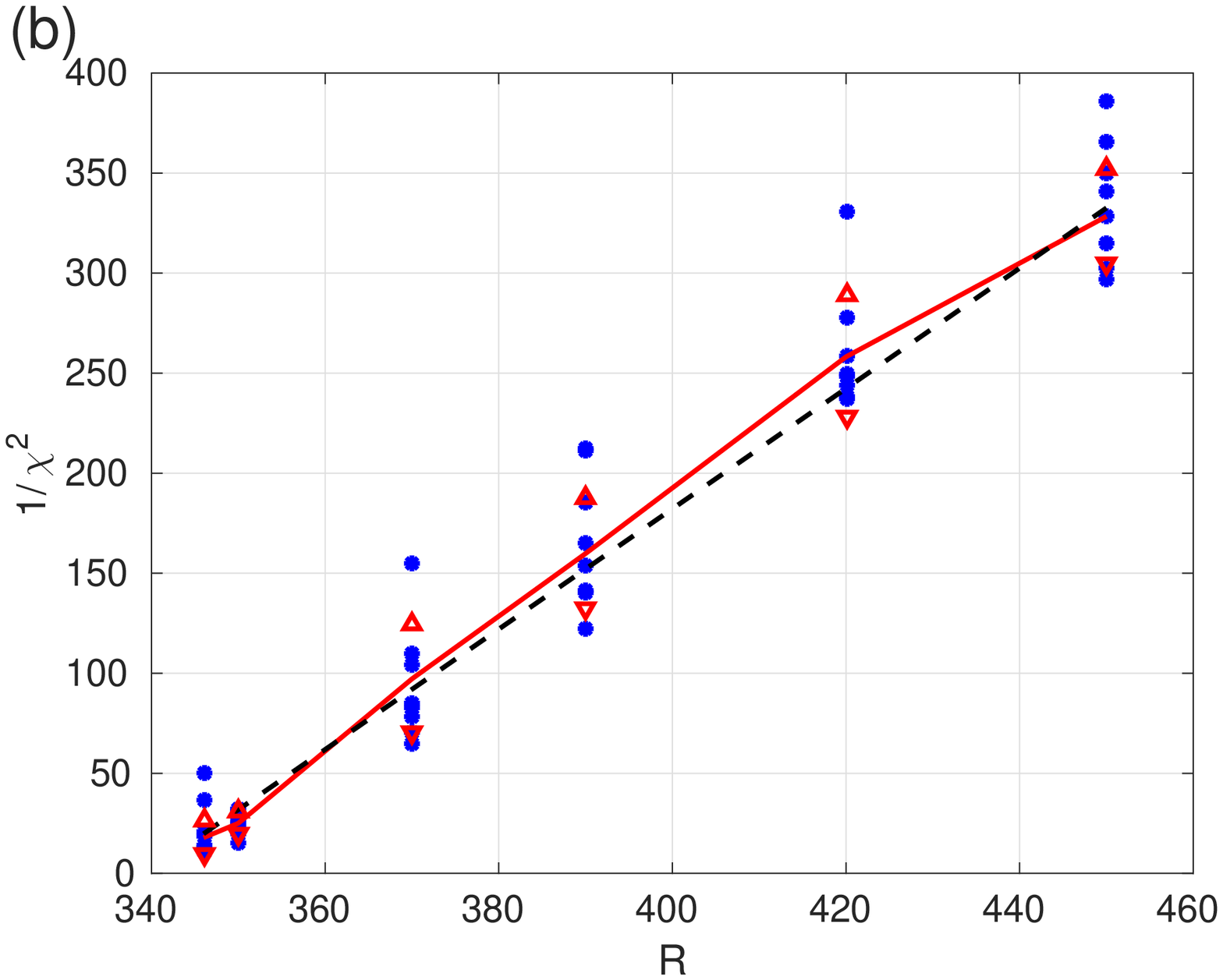}}
\caption{(a): Inverse of the square of the normalised order parameter as a function of Reynolds number in the uniform turbulence phase.
(b): inverse of the square of the response function as a function of the Reynolds number in the uniform turbulence phase.
The blue dots correspond to value at each given domain size and Reynolds number. The line corresponds to the average over
all domain sizes and the error bars correspond to the variance over all domain sizes.}
\label{mfs}
\end{figure}

Eventually, we consider the order parameter $M$ and its response function $\chi_M$ in the uniform turbulence phase, for $R$ above the $R_{\rm t}$ range.
Using the mean field description of the bands, we propose scaling laws in Reynolds number and size for these
two quantities $M\propto 1/\sqrt{L_xL_z|R-R_{\rm t,M}|}$ and $\chi_M \propto 1/\sqrt{|R-R_{\rm t,\chi}|}$ (\S~\ref{scmc}), where $R_{\rm t,M}$ and $R_{\rm t,\chi}$ are two fitting parameters falling in the $R_{\rm t}$ range. In this model, $A$ represents $m$, $\langle A\rangle$ represents $M$ and $\chi$ represents $\chi_M$. Within this framework,
the ratio $\chi_M/M$ also indicates us how shifted from $0$ the maximum of the PDF of $m$ is. In order to confront these
predictions to numerical results, we compute $1/(L_xL_zM^2)$ and $1/\chi^2$ in the low order simulations,
for each domain size, for $R\ge 346$. The values for the inverse of the order parameter are displayed as a
function of the Reynolds number in figure~\ref{mfs} (a), those for the inverse of the response function are displayed
in figure~\ref{mfs} (b). On top of the values for each domain size, we plot the average over all domain sizes.
The variance over all domain sizes provides the arrows. This firstly shows us that the scaling derived for the order
parameter is in very good agreement with numerical results. There is little dispersion about the average at each
Reynolds number, showing that $1/(M^2L_xL_z)$ depends very little on size. Moreover, we can see that this is an
affine function of the Reynolds number: the scaling in $R$ is also valid. We then consider $1/\chi_M^2$ in the same manner.
 The findings are very similar to what was found for $1/(L_xL_zM^2)$. At each Reynolds number,
 there is not much dispersion about the average. This confirms what could already be seen in other
 displays of response functions (Figure~\ref{fig2} (b), Figure~\ref{figkin} (c), Figure~\ref{KE} (c) and Figure~\ref{mchi} (b)):
 outside of the transitional range of Reynolds number, $\chi_M$ depends very little on size. Moreover, we can see that
 $1/\chi_M^2$ is very close to an affine function, thus verifying the predicted scaling in Reynolds number. Note that the fit of $1/\chi_M^2$ and $1/(L_xL_zM^2)$ by affine functions
 cross zero at slightly different Reynolds number $R_{\rm t,M}=340$ (for $M$) and $R_{\rm t,\chi}=327$ (for $\chi_M$). Note that these two Reynolds numbers have less physical relevance than $R_{c,1,2}$, so that we do not stress on them. We find a ratio of $\sigma/M\simeq 0.21$, this means that the parameter shifting the pdf maximum, when rescaled by size, noise amplitude, \emph{etc.} is of order $3$ (\S~\ref{scmc}, Eq.~\ref{rat}).

\section{Discussion}\label{concl}

The discussion will follow the line of statistical physics and phase transitions. In order to make things clear, we first remind that the two phases we consider
are uniform turbulence and oblique laminar-turbulent bands. We consider uniform turbulence as the disordered phase, at higher Reynolds number, and the banded phase as
the ordered phase, at low (though higher than $R_{\rm g}$) Reynolds number. Note that in that phase, there is a symmetry breaking, since an orientation is chosen (among two)
for the bands.

We first discuss the behaviour of the order parameter and the modulation of turbulence in bands.
Indeed, of both sampled quantities $M$ and $E$, the order parameter had the most typical behaviour during the crossovers:
that of a critical phenomenon\footnote{Invoking a critical phenomenon in the context of
clearly out of equilibrium forced dissipated transitional turbulence is not problematic, since this
is a concept which can be defined in both equilibrium states and Non Equilibrium Steady States. Indeed, this is
based on the non analyticity of the partition function \cite{jzj}. While many features of the bands can solely be described by a potential model, part of the wavenumber dependence of the modulation of turbulence is non potential \cite{RM10_1}.}.
We confirmed what had already been observed. Away from the
crossover, the order parameter has a concave decrease as the Reynolds number is increased, consistent with a square root scaling law. In this text, we could complement this
mean field picture by showing that both $M$ and $\chi_M$ had the inverse square root decrease with Reynolds number and size in the uniform turbulence
phase. Analysing a Ginzburg--Landau model for the crossover (\S~\ref{mfm}), an approach based on symmetries and motivated by the phenomenology of phase transitions,
yielded these disordered phase scalings in the context of a pdf maximum which is shifted (\S~\ref{scmc}). Including this shift of maximum explains
why the ratio of $M$ over its fluctuations is smaller than what would be expected in the simplest mean field model. Both the inclusion of this shift
and the ratio of $M$ over its fluctuations are in agreement with the data sampled in numerical simulations. We derived a Ginzburg criterion in our context (\S~\ref{gcrit})
to stress on the fact that these scalings are only valid away from the $R_{\rm t}$ range. Note that we resorted to the phenomenology of phase transitions
to derive results and analyse our data, since no first principles theory can be developed in forced-dissipated three dimensional plane Couette flow. This
is thus unlike two dimensional Euler-flows \cite{fetal,BSsns}. While being phenomenological, our approach still has more predictive abilities than $\mathbb{P}$-bifurcation
approaches used before in the data processing \cite{BT11}. The most striking of these predictions was that the match of the Landau model
in smaller size systems meant that a critical phenomenon could be expected in the larger systems. Performing the finite size analysis
confirmed that this prediction was correct. The decrease of $M$ and the maximum of $\chi_M$ linked to strong fluctuations, sampled in a
domain of a single size, could correspond to several scenarii.
However, the monotonous almost linear increase of the maximum of $\chi_M$ with size is the typical marker of a second order phase transition. This description can also lead to the proposition of
a first definition of critical Reynolds number: $R_{\rm c,1}$, the Reynolds number at which the maximum of $\chi_M$ is reached in an infinite size domain. This value falls in the $R_{\rm t}$ range. This threshold Reynolds number could be used in all two dimensional flows.

Following the kinetic energy of turbulence shows that the picture is actually more complex than that.
The first crossover leading to the decrease of modulation of turbulence and to orientation fluctuations of the bands is followed by a second crossover,
which contains reentrance of turbulence. While multistability appears (at least in small domains) between uniform turbulence and bands, and can seemingly
be recorded in time series of turbulent fraction or kinetic energy, the peculiar phenomenology leaves no traces in the PDF of $E$ (or turbulent fraction).
This makes the crossover undetectable by $\mathbb{P}$-bifurcation approaches. However, a crossover does happen and leaves clear traces in the detrended kinetic energy $E_{\rm det}$ which has a quick jump and in the response function
$\chi_E$ which has a maximum. This can lead to the proposition of a second threshold $R_{\rm c,2}$, the Reynolds number at which the asymptotic maximum of $\chi_E$ is reached, for the second crossover. Again, this falls within the $R_{\rm t}$ range. Performing the finite size analysis helps uncovering the nature of this crossover. A study
using a single size, showing the spatial and temporal coexistence of the two phases (bands and uniform turbulence), the fast increase of $E$ and the peak of $\chi_E$,
 could indeed lead one to think that one has a first order transition in an infinite size domain, with a discontinuity of $E$ (and $E_{\rm det}$) and a specific
 divergence of $\chi_E$ (see \cite{BL} and \S~\ref{rfo}). The monomodality of pdfs of $E$ shows that this is not the case. The finite size analysis confirms this
 by showing that $E_{\rm det}$ does not steepen past a certain size and the maximum of $\chi_E$ reaches an asymptotic value. All this phenomenology (the smoothness of
 $E_{\rm det}$, the convergence of $\max \chi_E$, the monomodality of pdf of $E$ and the spatial coexistence of both phase in quite small domains separated by fronts)
 is actually very similar to a smeared first order phase transition analysed in \ref{fstod}. In the model of this appendix, the banded phase can be represented by the $+1$ values of the rescaled field and the uniform turbulent phase can be represented by the $-1$ values of the rescaled field. The key ingredient of this scenario is that
 there is always a finite density of fronts between the two phases near the transition which prevents from performing the simple spacial averages
 used in the mean field study of first order phase transitions (\S~\ref{rfo}). This finite front density arises from a balance between some
 energy cost of having a front and the entropy from the disorder at finite noise amplitude. A similar argument leads to the finite density of defects in solids for instance (see \cite{kit} \S~20.1 for a rapid overview and \cite{ashmer} \S~30.1 for more details).
 Numerical simulations of a field model containing
 the smeared first order phase transition confirms that the kinetic energy of turbulence has all the features of this scenario.
 Moreover, the field model can be studied analytically right at the transition. The analysis is based on describing the field by a corresponding
 gas of non-interacting fronts, a method that had been introduced to show for instance that there exists non long range order in some one dimension
 systems. The method has recently been rediscovered in the study of multistability \cite{GZ,wpre}. The probability of having each front density can
 be derived in a finite size system. A two stages averaging procedure (first over front position at a front density then over front density) leads to
 the variance and the response function as a function of size right at the transition.  In the infinite size limit, one finds a simple equivalent of
 the response function, which is independent on size. This shows the origin of its finite asymptotic value (\S~\ref{dir},~\ref{per}).

Comparable studies can be performed in laboratory experiments, direct numerical simulations or low order simulations
of Couette flows, channel flows or pipe flows. One could verify if such a two stage scenario, implying two order parameters, like the one seen in two dimensional had disks systems \cite{hd,melting_hd}, is also found. The methodology imported from statistical physics and phase transitions could help make sense and propose
data processing in these future analyses. Among other things, this formalism indicates that there is always phase ordering dynamics, in a form
comparable to coarsening, which are longer and longer as domain size is increased. This means that a band orientation is always selected, even if this can
be extremely long in very large domains. Note that a finite density of front between domains of different orientations may appear in very large domains (see \cite{CTBtrans}). However,  the actual density
may be much lower than what may be suggested by relatively short simulations in large domains. Indeed, the description using fronts states that this density decreases exponentially fast with the ``potential'' cost of having a front. This cost is increasing as the Reynolds number is decreased and it quickly reaches a high enough value for the front density to be zero for all conceivable domain sizes. This gives all the more reasons to consider the phase ordering dynamics and their durations (however long) carefully, domain size by domain size: one cannot conclude on the question using one simulation of duration $\mathcal{O}(10^4)$ with one domain size. Note that other large scale effects may affect the bands ordering, such as soft Golstone modes leading to a slow turning of the band phase \cite{jzj}. Further studies could also help identifying the size scaling in quantities such as the Reynolds number at which the maximum of the response
function is reached. However, these studies may be tremendously expensive, due to extremely long time series necessary to see through the very strong fluctuations
occurring near $R_{\rm t}$. Simulation at an even lower order than the one used here may be necessary \cite{CTB,CTBtrans}. From another point of view, this provides another manner in which statistical physics, at equilibrium or out of equilibrium can be used to
study turbulence in its various forms.

\ack

The author thanks the hospitality of the Laboratoire d'Hydrodynamique de l'\'Ecole Polytechnique where this work was initiated, as well as that of Institut de Physique de Nice and Institute for atmospheric and Environmental sciences of Frankfurt, where the author worked when most of the research was conducted. This work was granted access to the HPC and visualization resources of ``Centre de Calcul Interactif'' hosted
by ``Universit\'e Nice Sophia Antipolis''.

\appendix \label{rap}

\section*{Appendix}

In the appendices, we will make great use of the phenomenology of phase transition to propose field models, based on the symmetries of the system,
that will describe very faithfully the physics of the transitional flow and predict precisely size and Reynolds number scalings (\cite{LL}).
In the two appendices, the model will be gradient, leading to a steady state which is at equilibrium, for simplicity of the early parts of the derivations.
Note that breaking this potential structure does not change much the derived scalings and the later parts of the derivations and the final results.

\section{Mean field and the phenomenology of phase transitions}\label{App}

In this first appendix, we will perform a mean field analysis in the uniform turbulence phase, to derive scalings for the order parameter and its response function. The mean field analysis,
requiring that the fluctuations of the order parameter are not too large will be given a bound of validity using a Ginzburg critertion.

\subsection{A mean field model}\label{mfm}

Let us consider only one orientation of the band and restart from the spatially averaged model of the modulus of the order parameter introduced in \cite{epje16}. This approach of modeling uses a Ginzburg--Landau equation. It is based on the symmetries of the system and its anisotropy. The choice of noise is based on numerical findings on the fluctuations of the amplitude of modulation of turbulence and of the kinetic energy. This model stands perfectly in the mindframe of the mean field phenomenology of phase transition. It is a potential model and reads
 \begin{equation}
 \tau_0 \frac{dA}{dt}=-\partial_A V+\zeta(t) \,, \,
 V=\frac{|\epsilon|}{2}(A-\Delta)^2+\frac{g}{4}A^4-\frac{\alpha^2}{2}\ln(A)\,,\, \langle \zeta(t)\zeta(t')\rangle=a^2\delta(t-t')\,.\label{eqchvar}
\end{equation}
The amplitude of the noise $\zeta$ is $a^2=\alpha^2/(L_xL_z)$, where $\alpha$ is the amplitude of the noise felt locally by $A$ and $L_x$ and $L_z$
are the streamwise and spanwise sizes of the domain. This scaling comes from the basic operation of the mean field analysis: a spatial average. Indeed, the variance of a noise white averaged in space in two dimension is one over the surface of average \cite{gar,VK}. Assuming that
the fluctuations of $A$ are not too large, we can replace the field by its spatial average. This will be valid in the uniform turbulence phase, due to the spatial homogeneity and absence of fronts between different phases. A similar analysis had been performed in the band phase, away from $R_{\rm t}$, again when the amplitude of the order parameter was fluctuating weakly around its average.
 In order to work in the disordered phase, we use the absolute value of the distance
to the threshold of transition $|\epsilon|$. Note that we introduced a time independent additive parameter $\Delta$ in order to account
for effects which can push $A$ further away from $0$ in the disordered phase. The factor $\tau_0$  gives a time scale and $g$ measures the amplitude
of non-linearities.

Starting from the overdamped Langevin equation~(\ref{eqchvar}), we write the Fokker--Plank equation for the dynamics of the pdf of $A$, $P(A,t)$. This reads
\begin{equation}
\frac{\partial P}{\partial t}=\frac{\partial~}{\partial A}\left(\frac{dV}{dA}P \right)+\frac{a^2}{2}\frac{\partial^2 P}{\partial A^2}\,.
\end{equation}
In the steady regime, we can solve this equation (without probably flux) and obtain the pdf of $A$ using the potential $V$.
\begin{equation}
P=\frac{1}{Z}\exp\left(-\frac{2L_xL_zV}{\alpha^2} \right)\,.
\end{equation}
The factor $Z$ is the partition function and normalises the pdf when no forcing field are applied on the system.

\subsection{Average and variance in disordered phase}\label{scmc}

Using this pdf, we can compute the average and the fluctuations in the disordered phase. We will neglect the effect of non linearities $g$
and focus on the scaling with $|\epsilon|$ and the size $L_xL_z$, as well as the effect of this force $\Delta$. We can thus write the pdf,
the partition function and the moments
\begin{eqnarray}
P=\frac{A\exp\left( -\frac{|\epsilon| L_xL_z(A-\Delta)}{\alpha^2} \right)}{Z}\,, \\ Z=\int_{A=0}^\infty A\exp\left( -\frac{2|\epsilon| L_x L_z(A-\Delta)}{\alpha^2} \right)\,{\rm d}A\,,\\ \langle A^n\rangle=\frac{1}{Z}\int_{A=0}^\infty A^{n+1}\exp\left( -\frac{2|\epsilon| L_x L_z(A-\Delta)}{\alpha^2} \right)\,{\rm d}A\label{pdf}
\end{eqnarray}
The average, the fluctuations and the ratio of average to fluctuations are calculated at first order in $\Delta$.
In order to simplify calculations, let us perform a rescaling of $A$:
$B=(A-\Delta)\sqrt{2|\epsilon| L_x L_z}/\alpha$ and of $\Delta$: $D=\Delta\sqrt{2|\epsilon| L_x L_z}/\alpha$. We also introduce the integrals $J_n$
which will be systematically calculated
\begin{equation}
J_n\equiv\int_{-D}^\infty (B+D)^n\exp\left( -\frac{B^2}{2}\right)\,{\rm d}B\,.
\end{equation}
They depend on $D$ alone. The normalisation $Z$ and the moments $\langle A^n\rangle$ can be simply written as a product of $J_n$ integrals
and powers of $\alpha/\sqrt{\epsilon L_x L_z}$.
\begin{equation}
Z=\left(\frac{\alpha}{\sqrt{\epsilon L_xL_z}}\right)^2J_1\, ,\, \langle A^n\rangle=\frac{1}{Z}\left(\frac{\alpha}{\sqrt{\epsilon L_xL_z}}\right)^{n+2}J_{n+1}=\left(\frac{\alpha}{\sqrt{\epsilon L_xL_z}}\right)^n\frac{J_{n+1}}{J_1}\,.
\end{equation}
Since $D$ is not so large, we will work at first order and expand the integrals $J_n$. This is physically justified and will help the calculation.
It is expected that $J_n$ will follow the linear tendency outside the range of validity of the expansion, even if it displays a curvature.
The expansion reads
\begin{equation}\eqalign{
J_n(D)=J_n(0)+D\frac{\partial J_n}{\partial D}(0)+\mathcal{O}(D^2)=\int_{B=0}^\infty B^n\exp\left( -\frac{B^2}{2}\right)\,{\rm d}B
 \cr +D\left(n\int_{B=0}^\infty B^{n-1}\exp\left( -\frac{B^2}{2}\right)\,{\rm d}B\right)+\mathcal{O}(D^2)\,.}
\end{equation}
One then has two cases
\begin{equation}\eqalign{
J_0(D)=\int_{B=0}^\infty\exp\left( -\frac{B^2}{2}\right)\,{\rm d}B +D\exp\left( -\frac{D^2}{2}\right)+\mathcal{O}(D^2) \,,
\cr J_{n>0}(D)=\int_{B=0}^\infty B^n\exp\left( -\frac{B^2}{2}\right)\,{\rm d}B+nD\int_{B=0}^\infty B^{n-1}\exp\left( -\frac{B^2}{2}\right)\,{\rm d}B+\mathcal{O}(D^2)\,.}
\end{equation}
At leading order, the variation of the bounds is the only $\mathcal{O}(D)$ contribution to the differentiation under the integral sign if $n=0$, while it is absent if $n>0$: the integrand is zero at both bounds.
These integrals can be generally rewritten using the integral $I_n(D)$.
\begin{equation}
I_n\equiv \int_{B=0}^\infty B^n\exp\left( -\frac{B^2}{2}\right)\,{\rm d}B\,.
\end{equation}
One can obtain a recurrence relation using an integration by part
$I_n=(n-1)I_{n-2}$. The zeroth integral is a gaussian integral $I_0=\sqrt{\frac{\pi}{2}}$ and the first integral is directly integrated $I_1=1$. The following integrals are obtained using the recurrence relation, so that one has $I_2=\sqrt{\frac{\pi}{2}}$ and $I_3=2$. This yields the required $J_n$
\begin{equation}
J_1(D)=1+D\sqrt{\frac{\pi}{2}}+\mathcal{O}(D^2)\,,\,
J_2(D)=\sqrt{\frac{\pi}{2}}+2D+\mathcal{O}(D^2)\,,\,
J_3(D)=2+3\sqrt{\frac{\pi}{2}}D+\mathcal{O}(D^2)\,.
\end{equation}
From there we obtain the cumulants
\begin{equation}
\langle A\rangle=\frac{\alpha}{\sqrt{2|\epsilon| L_xL_z}}\frac{J_2}{J_1}
=\frac{\alpha}{\sqrt{2|\epsilon| L_xL_z}}\left( \sqrt{\frac{\pi}{2}}+D\left(2-\frac{\pi}{2} \right)+\mathcal{O}(D^2)\right)\label{av_ap}\,,
\end{equation}
\begin{equation}
\eqalign{
\sigma^2=\langle A^2\rangle -\langle A\rangle^2=\frac{\alpha^2}{2|\epsilon| L_xL_z}\left(\frac{J_3J_1-J_2^2}{J_1^2}\right)=\cr \frac{\alpha^2}{2|\epsilon| L_xL_z}\left(2-\frac{\pi}{2}+D\sqrt{\frac{\pi}{2}}(\pi-3)+\mathcal{O}(D^2) \right)\,,} \label{var_ap}\end{equation}
\begin{equation}
\frac{\sigma^2}{\langle A\rangle^2}=\left(\frac{J_3J_1}{J_2^2}-1\right)=\frac{4}{\pi}-1+D\underbrace{\left(5\sqrt{\frac{2}{\pi}}-8\left( \frac{2}{\pi}\right)^{\frac{3}{2}}\right)}_{\simeq -0.07<0}+\mathcal{O}(D^2)\,. \label{rat_ap}
\end{equation}
 The results read
\begin{eqnarray}
\langle A\rangle \label{av}
=\frac{\alpha}{\sqrt{2|\epsilon| L_xL_z}} \sqrt{\frac{\pi}{2}}+\Delta\left(2-\frac{\pi}{2} \right)+\mathcal{O}(\Delta^2)\,,\\
\sigma^2=\frac{\alpha^2}{2|\epsilon| L_xL_z}\left(2-\frac{\pi}{2}\right)+\Delta\sqrt{\frac{\pi}{2}}(\pi-3)+\mathcal{O}(\Delta^2) \,, \label{var}
\end{eqnarray}
\begin{equation}
\frac{\sigma^2}{\langle A\rangle^2}=\underbrace{\frac{4}{\pi}-1}_{\simeq 0.27}+\frac{\sqrt{2|\epsilon| L_xL_z}}{\alpha}\Delta\underbrace{\left(5\sqrt{\frac{2}{\pi}}-8\left( \frac{2}{\pi}\right)^{\frac{3}{2}}\right)}_{\simeq -0.07<0}+\mathcal{O}\left(\frac{|\epsilon| L_xL_z}{\alpha^2}\Delta^2\right)\,. \label{rat}
\end{equation}
Let us first consider only the average $\langle A\rangle$ and fluctuations $\sigma$ in the limit $\Delta=0$. They both have the same scaling in $1/\sqrt{|\epsilon L_xL_z}$. This shows us that both average and fluctuations decrease with the distance to the transition threshold and with the size. They are zero in the limit of infinite domain size. Note however that the response function $\chi=\sqrt{L_xL_z}\sigma$ is always finite. This scaling can be easily identified from numerical or experimental data. Indeed, in that case $1/(L_xL_z\langle A\rangle^2)$ and $1/\chi^2$ are both linear in the distance to the threshold $|\epsilon|$. The effect of the force $\Delta$, which also shifts the maximum of the pdf $P(A)$ further from zero can easily be seen in the ratio of variance to average. The variance is smaller than the average: increasing $\Delta$ makes this ratio even smaller. The reason is simple: when $\Delta$ is increased, the average is also significantly increased, with a factor $\simeq 0.5$. This parameter impacts very little the shape of the pdf, so that the fluctuations are increased at a smaller rate $\simeq 1.8$. As a consequence, the ratio decreases slowly with $\Delta$. Another information can be obtained from this ratio: if it is independent on size or $\epsilon$, then the shift $\Delta$ scales like $\alpha/\sqrt{|\epsilon|L_xL_z}$. As a consequence, the introduction of a non zero $\Delta$ would not impact the scaling of $\langle A\rangle$ and $\sigma$ nor would it impact the manner of identifying this scaling.

\subsection{Ginzburg Criterion}\label{gcrit}

The derivation of the former section, or the scaling of $\langle A\rangle$ in $\epsilon$ in the ordered phase \cite{RM10_1,prigent02,pr_physD} are valid if relative fluctuations are small and if the corresponding non linearities in the mean field models can be neglected. In the disordered phase, this corresponds to the $gA^4$ term in equation~(\ref{eqchvar}), while the ordered phase, this corresponds to the quartic term coupling the two orientations. One can estimate a range of parameter in which the mean field scalings are valid and a range of parameters in which the scalings of $\langle A\rangle$ and $\chi$ in $\epsilon$ and size $L_xL_z$ are more complex. In that case, one has to compute them using renormalisation group methods (in the thermodynamic limit) or numerical simulations (in finite size domains).

A quantitative manner of estimating the range of validity of the mean field approximation is to compute the corresponds brought to the partition function by the non-linearities. One obtains a \emph{Ginzburg criterion} on the distance to the threshold. This calculation is simple in the disordered phase \cite{LL}. It is longer in the ordered phase where the coupling of orientation has to be taken into account.
A Ginzburg type criterion can be derived to discuss the range of validity of mean field calculation of the former section. It amounts to taking into account the quartic terms in a perturbative manner at first order, for instance in the computation of the partition function. We then determine in which limit they can be neglected. We use for instance the spatial independent, quartic potential with one orientation:
\begin{equation}
V\simeq\int_{0}^{L_x,L_z} {\rm d}x{\rm d}z\,\frac{|\epsilon|}{2}A^2+\frac{g_1}{4}A^4 \,.
\end{equation}
The partition function is:
\begin{equation} Z=\int {\rm d}A\, A \exp\left(-\frac{2}{\alpha^2} L_xL_z\left(\frac{|\epsilon|}{^2}A^2+\frac{g}{4}A^4\right)\right)\,.
\end{equation}
The exponential $\exp(-2(L_xL_z)/\alpha^2 (gA^4/4))$ is developed at first order, so that the partition function reads
\begin{equation}
Z=\frac{\alpha^2}{2L_xL_z|\epsilon|}I_1 -\frac{g\alpha^4}{16|\epsilon^3|(L_xL_z)^2}I_5=\frac{\alpha^2}{2L_xL_z|\epsilon|} -\frac{g\alpha^4}{2|\epsilon^3|(L_xL_z)^2}\,,
\end{equation}
with $I_5=4I_3=8$. We can see that the quartic term is negligible if:
\begin{equation}
\frac{\alpha^2g}{\epsilon^2L_xL_z} <1\Leftrightarrow \frac{\sigma\sqrt{2/(2-\frac{\pi}{2})}}{\sqrt{|\epsilon|/g}}<1\,.
\end{equation}
This inequality on $|\epsilon|$ is the Ginzburg criterion

In practice, this corresponds to assuming that the fluctuations are small relatively to the order phase mean field average value of $A_\pm$, which is a classical result \cite{LL,jzj}. This indicates that the scaling laws should be used away from the transition. Besides, the size of the non-mean field range decreases with the size. Note that it is a finite size scaling law, since it involves $L_xL_z$. In the thermodynamic limit, the scaling law is along the lines of $\chi^2/(A^2\zeta^2)<1$.

\section{Smeared first order phase transitions}\label{fstod}

In this appendix, we present a model displaying a smeared first order phase transition, in order to illustrate the scenario observed in the
kinetic energy of the flow. We first remind the typical model of a first order phase transition, the necessary hypotheses and the subsequent scaling (\S~\ref{rfo}).
We then present the principle of the smeared first order transition and explain how the presence of finite density of fronts between the two phases change the scenario (\S~\ref{princ}).
We first illustrate this by numerical simulations of a model displaying such a transition (\S~\ref{simusmear}), then derive the scaling of the response function at the transition (\S~\ref{scalsmear}).

\subsection{Finite size analysis of a classical first order transition }\label{rfo}

First order phase transitions are probably the most widely known type of phase transitions, because one encounters them regularly in its everyday life. Let us set $\epsilon$ the parameter which controls the system, $L$ the size in each of the dimension $d$ \cite{LL}. Let us consider how the thermodynamic limit is reached in a first order transition, and show the scalings followed by the average, variance and pdf using the asymmetric Allen--Cahn equation
\begin{equation}\eqalign{
\frac{\partial A}{\partial t}=\frac{\delta V}{\delta A}+\sqrt{\frac{2}{\beta}}\eta(\vec x,t)\,, \cr 0\le x_{1\le i\le d}\le L  \,,\, \langle \eta(x,t)\eta(x',t')\rangle=\delta(x-x')\delta (t-t')\,,\cr V=\int_{0}^{L}\left(\epsilon A-\frac{A^2}{2}+\frac{A^4}{4}+\frac{1}{2}\sum_{i=1}^d\left( \frac{\partial A}{\partial x_i} \right)^2\right)\,.}\label{mod}
\end{equation}
For convenience, this model is potential and uses white noise: all the properties we derive here can be transposed to most non potential correlated and coloured noise, using more technical derivations. The main results will not be changed is one uses periodic or Dirichlet $A=0$ boundary conditions. One of the interesting property of such a system is that one can write the equilibrium probability density function of $A$ in the form
\begin{equation}
\rho=\frac{\exp(-\beta V)}{\int \exp(-\beta V)\,{\mathcal{D}A}} \,,\label{pdfallencahn}
\end{equation}
without much technicalities. This expression is deduced from the Fokker--Plank equation for $\rho$ which is equivalent to the overdamped Langevin equation~\ref{mod} for $A$ (see \cite{gar} for a review both type of equations and \cite{RM10_1,epje16} for a example of the use in a model of transition). A comparable writing is possible in non potential systems, but it is slightly more complex and does not bring additional insight to our discussion.
The first operation of the mean field description of the phase transition in this model is to consider the spatial average of the field $\bar{A}\equiv (1/L^d)\int_{x_i=1}^L A \,{\rm d}^dx $. We will always consider spatial averages because it indicates us whether there is order. If one has $A(x,t)\simeq \bar{A}(t)$ ``nearly everywhere'' and ``most of the time'', then one can spatially average the evolution equation or simply integrate the potential $V$, which yields $V\simeq L^d(\epsilon \bar{A}-A^2/2+A^4/4)$. One expects this to be valid if the strength of the noise felt. The precise criterion of validity of this approximation will be presented and discussed in the next section. In that regime of approximation, processing the pdf Eq.~(\ref{pdfallencahn}) is fairly simple and is often done in the literature \cite{BL}. One uses the mean field approximation: the probability density function is approximated by a sum of gaussian centered on the minima of $V$ whose variance is given by the second derivatives of $V$. Provided that $|\epsilon|<2/(3\sqrt{3})$, there are two minima, one positive and one negative. Moreover, if $\epsilon$ is small, one can even obtain an expression at first order in $\epsilon$ of the minima of $V$, of its second derivatives and of the values of $\bar{A}$ at which the minima are reached.

In order to be in the regime of study of the phase transition in the large size limit, one can state the approximation more precisely as $|\epsilon|\ll \beta L |\epsilon| \ll 1$, one then has a simple expression for
 the probability density function
\begin{equation}\label{pdfsymp}
\eqalign{
\rho=\frac{1}{Y}\left( \exp(-\beta L^d  \epsilon)\exp(-\beta L^d(\bar{A}-1)^2) +\exp(\beta L^d \epsilon)\exp(-\beta L^d(\bar{A}+1)^2)\right)
 \,, \cr  Y=2\cosh (\beta L^d  \epsilon)\sqrt{\frac{\pi}{\beta L^d}}\,.}
\end{equation}
We find bimodality in the probability density function, in particular when $\epsilon=0$, where the pdf is exactly symmetric. In that case, it tends toward two Diracs in the infinite size limit. When $\epsilon\ne 0$, the pdf is asymmetric: observing $\bar{A}=\pm1$ is more probable if $\epsilon=\mp 1$. Any field can be turned in such a $A$ by a shift of the average and a rescaling. The pdf is all the more asymmetric that the size is increased, as shown by the exponential factors: in the thermodynamic limit one observes only one phase if $\epsilon\ne 0$. This of course impacts the the ensemble average of $\bar{A}$
\begin{equation}
\langle \bar{A}\rangle=-\tanh(\beta L^d \epsilon)\,,\label{mfaca}
\end{equation}
Which goes from $\langle \bar{A}\rangle=1$ if $\epsilon \rightarrow -\infty$ to $\langle \bar{A}\rangle=-1$ if $\epsilon\rightarrow+\infty$. This change is smooth at finite size, but it is faster and faster as size is increased
as shown by the increasing slope at $\epsilon=0$. In the thermodynamic limit, the average is discontinuous at $\epsilon=0$. This also gives us the response function
\begin{equation}
\chi^2=L^d\left(\langle \bar{A}^2\rangle-\langle \bar{A}\rangle^2\right)=\frac{1}{2\beta }+\frac{L^d}{\cosh^2(\beta L^d\epsilon)}\,.\label{cmf}
\end{equation}
The inverse of the square hyperbolic cosine will decrease very fast is the argument is away from $0$. This means that outside a band of $\delta\epsilon \simeq 1/(L^d\beta)$, the response function is finite and size independent, and in fact given by the noise amplitude. Things are different in the band $\delta\epsilon \simeq 1/(L^d\beta)$ around $\epsilon=0$, where the hyperbolic cosine is close to one. One then find a peak in the response function whose amplitude grows like $L^d$. This means that at the transition, $\epsilon=0$, the response function will diverge in the thermodynamic limit. Note however that this divergence is different from the power law divergences of critical phenomena $|\epsilon|^\nu$, because the hyperbolic cosine will remove all increase of $\chi$ with $L$ if $\epsilon \ne 0$. The origin of this particular divergence is that the variance of $A$ is of order one at the transition, since in the large size limit, one has $\bar{A}=\pm1$ each with probability one half at $\epsilon=1$, while one has $A=1$ or $A=-1$ with probability one at $\epsilon\ne 0$ (so that variance is zero). Taking the response function thus leads to the divergence right at $\epsilon=0$.

The calculations leading to these results are very similar in their spirit to those performed in appendix~\ref{App}.

\subsection{Finite size analysis of a smeared first order transition }\label{princ}

We now present the transition of interest and first explain what constraint we release in the former analysis: we assume that $\beta$ is not so large that noise fluctuations can naturally create fronts between the two phases
with a non negligible probability. This phenomenon has already been seen: this destroys long range order in one dimensional spin systems. This approach has also been used recently to
describe transitions between $A=1$ everywhere and $A=-1$ everywhere in Allen-Cahn system outside of the usual low noise limit \cite{GZ}. Since there will be
at least one front in the system most of the time, the powers of the spatial averages will certainly not be equal to the spatial averages of the powers in equation~(\ref{mod}). Moreover, the gradient terms will be non zero and also impact the potential. This can strongly modify the pdf and the cumulants of $A$ and in particular erase the divergence of the response function right at the transition. We will illustrate how and why, first by numerical simulations of equation~(\ref{mod}), then by an analytical analysis.

\subsection{Simulations of the smeared transition}\label{simusmear}

\begin{figure}
\centerline{\includegraphics[width=15cm]{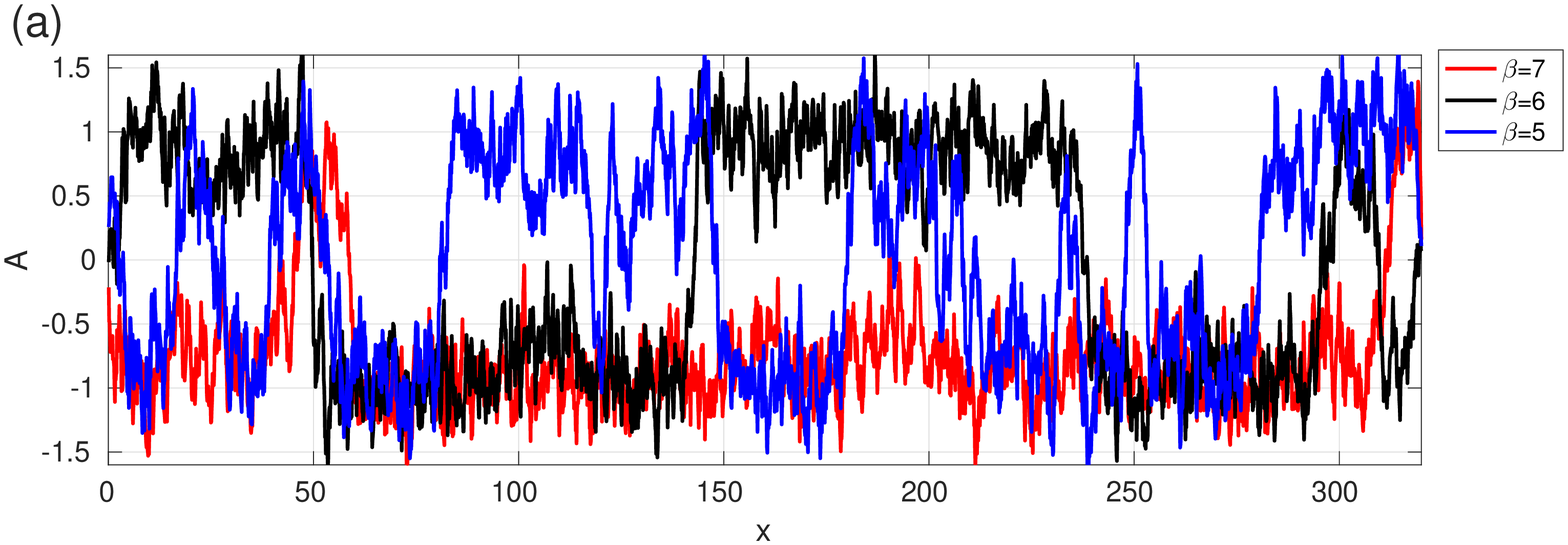}}
\centerline{\includegraphics[width=6cm]{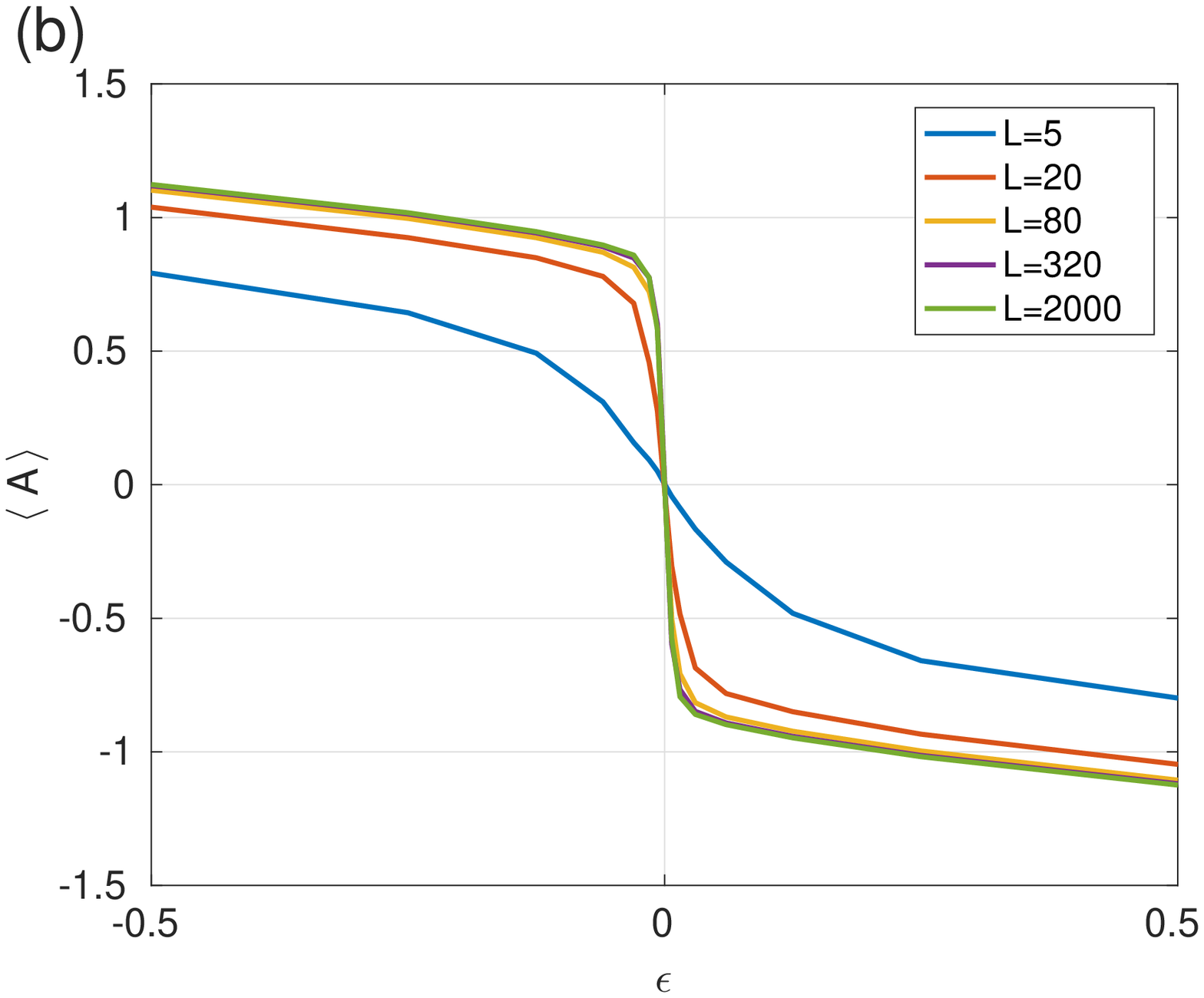}\includegraphics[width=6cm]{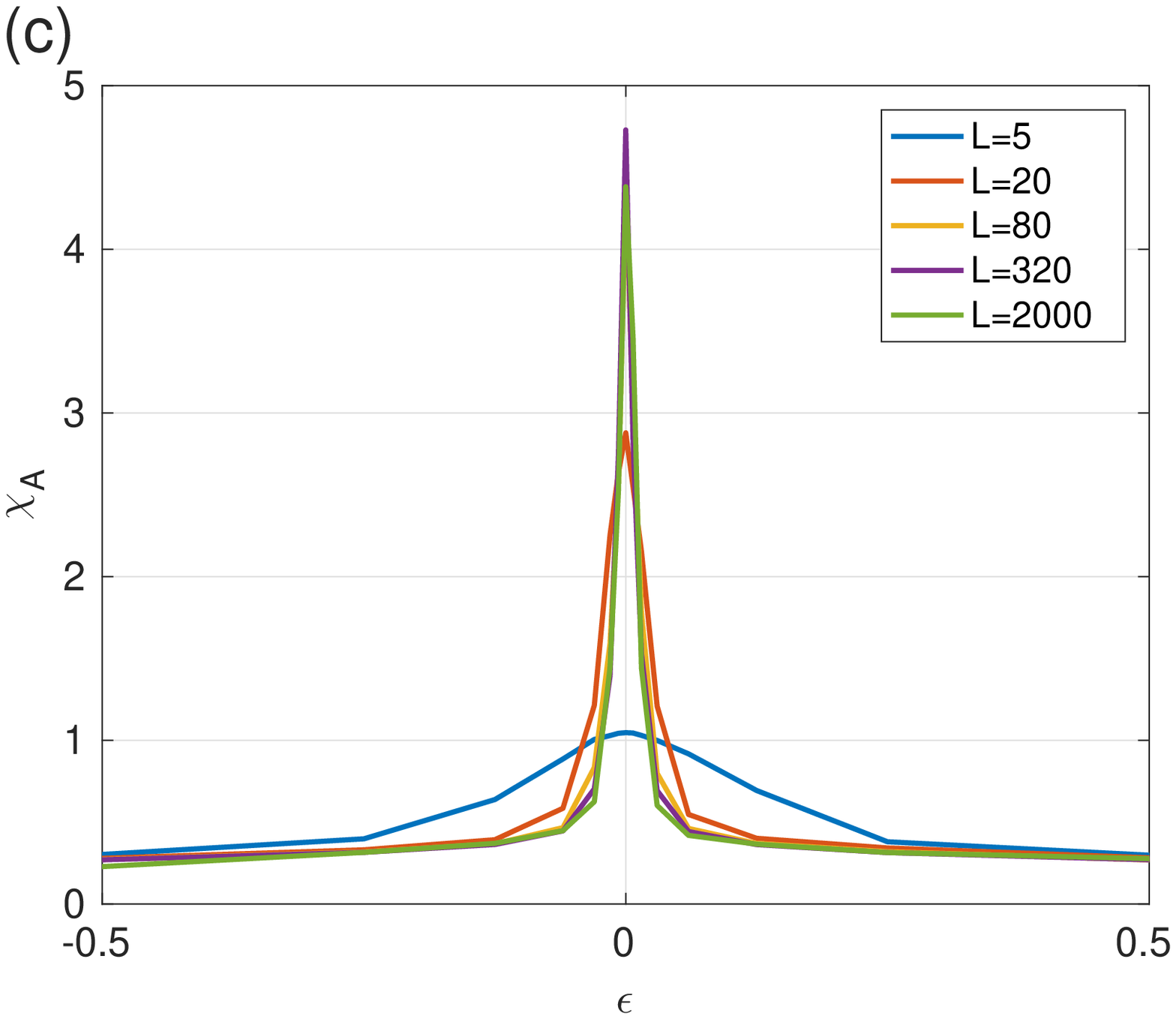}}
\centerline{\includegraphics[width=6cm]{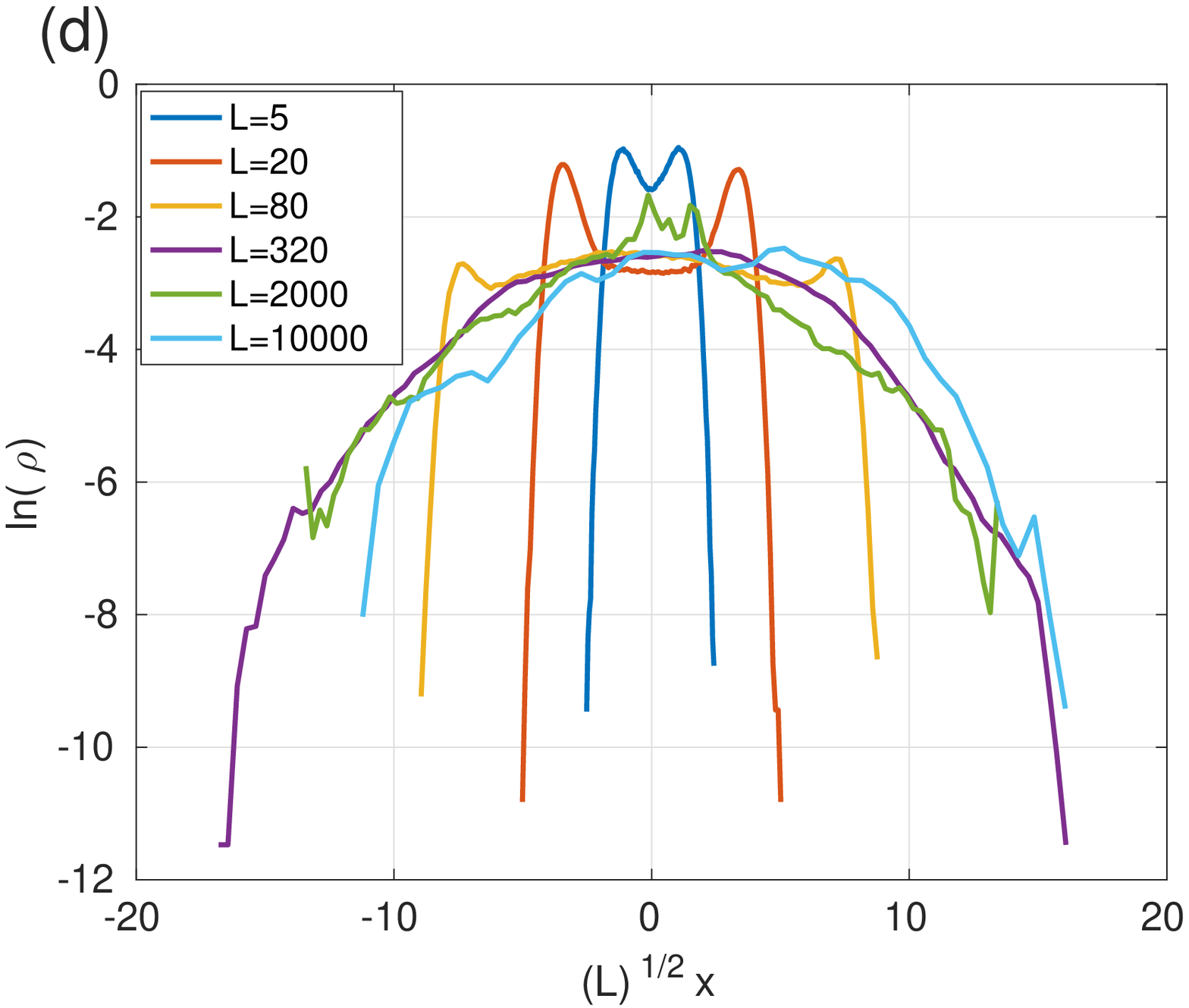}\includegraphics[width=6cm]{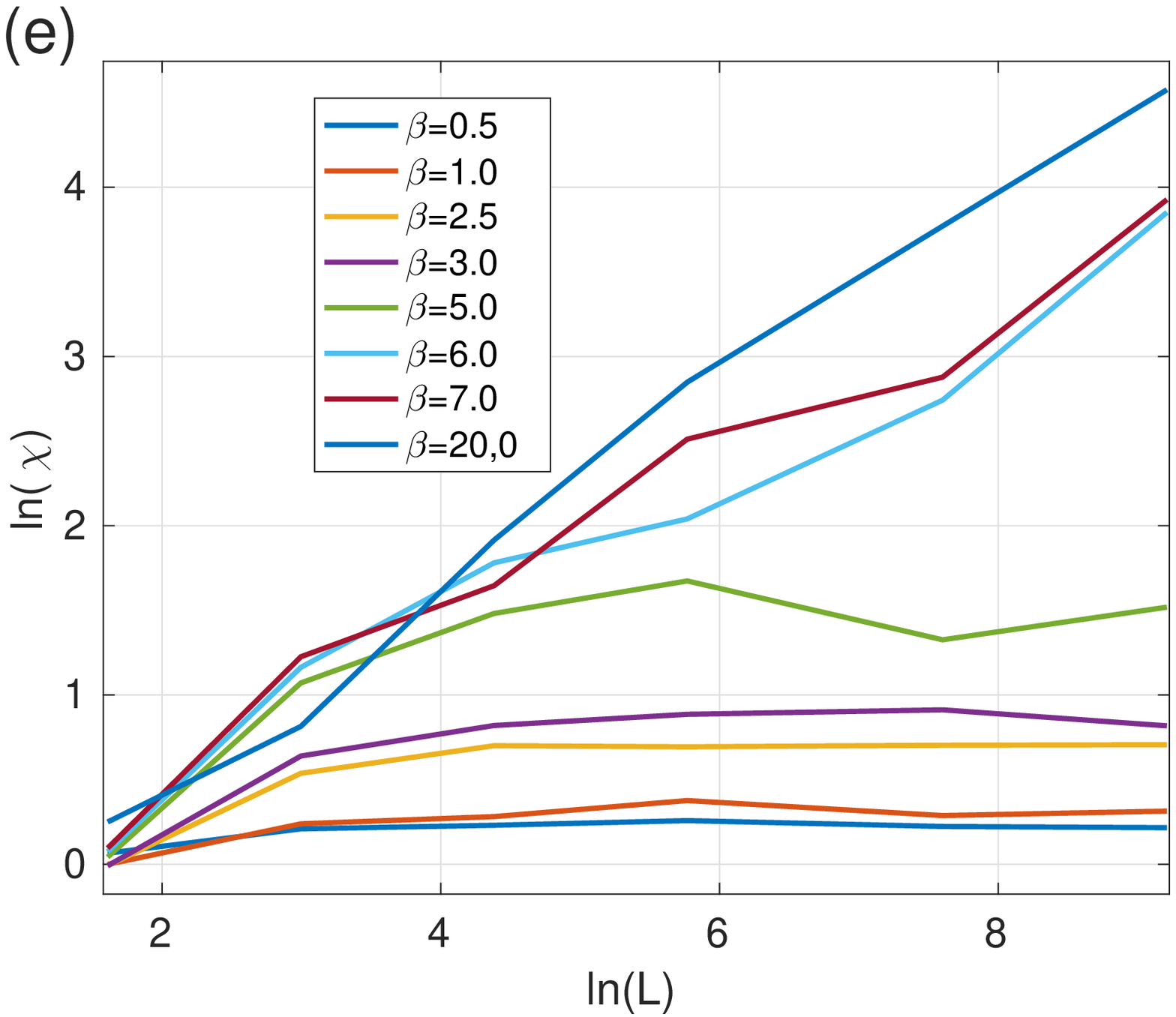}}
\centerline{\includegraphics[width=6cm]{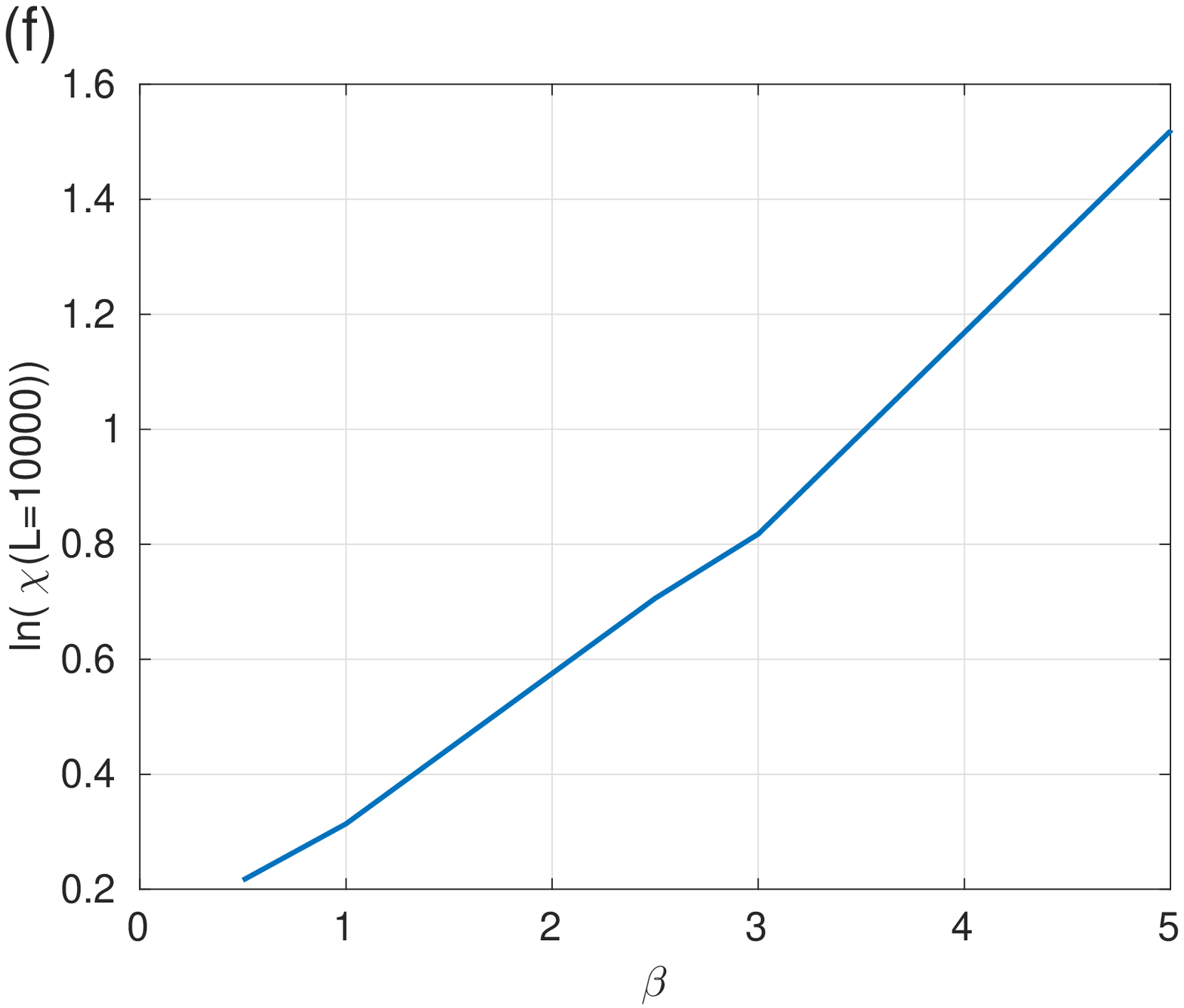}}
\caption{(a) Instantaneous view of the field $A$ obtained from numerical integration of the Asymmetric Allen-Cahn equation right at the transition $\epsilon=0$ as a function of space in a domain of size $L=320$ for increasing values of $\beta$. (b) Ensemble average of the spatial average $\langle \bar{A}\rangle$ of the field as a function of $\epsilon$. (c) response function of the field $\chi_A$. (d) logarithm of the probability density function of the spatially averaged field $\bar{A}\sqrt{L}$ normalised by length. (e) Logarithm of the maximum of the response function as a function of the logarithm of the size for increasing values of $\beta$. (f) Logarithm of the response function at $l=10000$ as a function of $\beta \in[0.5; 5]$}
\label{simusAC}
\end{figure}

We simulate the stochastic partial differential equation~(\ref{mod}). Finite differences are used in space and a semi-implicit Euler algorithm is used to advance the equation in time. We use $dx=$ and $dt=$. We will save instantaneous fields $A(x)$ (Figure~\ref{simusAC} (a)) and systematically sample the spatial average $\bar{A}$ to compute probability density functions (Figure~\ref{simusAC} (d)), ensemble averages (Figure~\ref{simusAC} (b)) and response functions (Figure~\ref{simusAC} (c,e,f)).

The spatial state of the field $A$ can be illustrated by instantaneous view in a domain of length $L=320$, simulated at three $\beta=5$, $\beta=6$ and $\beta=7$ at $\epsilon=0$, thus forced by weaker and weaker noise (Figure~\ref{simusAC} (a)).
The spatial variations of the field are of two kinds: small fluctuations either around $+1$ or $-1$ or rapid jump between the neighbourhood of $+1$ to that of $-1$. These later variations are the fronts between domain of phase $+1$ and domains of phase $-1$. Note that as $\beta$ is increased and the forcing noise amplitude is decreased, the front density is severely decreased and the typical length of domains of a given phase is severely increased. For an even larger $\beta=20$, no fronts where seen even in the largest domains of length $L=10000$ which were simulated. This hints that as $\beta$ is decreased, the system is less and less likely to be well described by its mean field approximation. For instance at $\beta=5$, $\epsilon=0$, the spatial average is always near $0$, while the absolute value of the field $|A|$ is always near $1$. We can have a more precise view of the effect of decreasing $\beta$ and thus destroying order by considering the $\epsilon$ dependence of the ensemble average $\langle \bar{A}\rangle$, in figure~\ref{simusAC} (b), computed from systems of increasing length at $\beta=5$. One can see that $\langle \bar{A}\rangle (\epsilon)$ converges toward an asymptotic behaviour as $L$ is increased. There is little differences between values sampled in a domain of sizes $L=80$ to $L=2000$. The other striking fact is that this asymptotic behaviour shows a rapid though continuous change from $\langle \bar{A}\rangle \simeq -1$ to $\langle \bar{A}\rangle\simeq +1$ at $\epsilon\simeq 0$, which is quite different form the mean field, possibly high $\beta$ behaviour (Eq.~(\ref{mfaca})), which becomes discontinuous as size go to infinity. This difference with the mean field behaviour can also be seen in the response function as a function of $\epsilon$ at $\beta=5$ (Figure~\ref{simusAC} (c)) sampled in domains of increasing sizes. Again, one can note convergence toward an infinite size behaviour, which presents a finite response function maximum at $\epsilon=0$ and peak of finite thickness around $\epsilon=0$. Again, this is in clear disagreement with the divergence of the mean field behaviour (Eq.~(\ref{cmf})). We eventually consider the probability density functions of $\bar{A}$ at $\epsilon=0$ and $\beta=5$ for increasing length (Figure~\ref{simusAC} (d)). We note that for smaller domains, the pdf are bimodal. However, this bimodality is weaker and weaker and completely disappears as size is increased. The pdf converge toward some bell shape of width given by $\sqrt{L}$. It appears that the bimodality is erased by the impossibility of observing a coherent phase over the whole domain. This is again quite different from what is seen in the mean field approximation (Eq.~(\ref{pdfsymp})), where the pdf tended toward two Diracs.

The degree to which the crossover is similar or different from the meanfield is actually controlled by $\beta$. This can be seen for instance by considering the maximum of the response function (always found at $\epsilon=0$) as a function of the domain size for increasing $\beta$ (Figure~\ref{simusAC} (e)). The maximum of the response function reaches a plateau as the size is increased. However, this plateau is reached later and later as $\beta$ is increased. This plateau value is larger and larger as $\beta$ is increased. This increase with $\beta$ appears to be close to exponential (Figure~\ref{simusAC}). We also note that at $\beta=20$, nothing distinguishes $\chi(\epsilon=0,L)$ from the mean field behaviour, which predicts $\chi(\epsilon=0,L)\simeq \sqrt{L}$.  The same goes for the pdfs at $\epsilon=0$, which remain bimodal and peaked around $\pm1$, and for the average of the field $\langle \bar{A}\rangle$ which indeed becomes discontinuous at $\epsilon=0$. We can thus draw the line between two behaviour. One the one hand we describe \emph{smeared} transition, which reach relatively quickly an asymptotic state as size is increased which is different from the mean field, with rapid though continuous average, movement between two phases with monomodal pdf and peaked though finite response function. One the other hand we describe the classical first order phase transitions which display the mean field features scalings for all conceivable domain sizes. The band of $\beta$ in which these \emph{smeared} transitions occur appears to be quite narrow, so that they have little relevance for most condensed matter systems.

\subsection{Analytics of the smeared transition}\label{scalsmear}

\subsubsection{Principle}

here, we calculate the average and the variance of the spatial average of the field $\bar{A}$ when the noise amplitude is large enough for several fronts between domains of $A=+1$ and $A=-1$ to exist in the system. We perform this at $\epsilon=0$, right at the transition. There, it is the easiest to discuss the scalings of the average and response function of the field and distinguish between the classical mean field case and the high noise, spatially inhomogeneous case. We perform the derivation for a system of size $L$ with one dimension of space. For this matter, we prolong the analysis performed in the study of metastability in the Ginzburg--Landau--Allen--Cahn equation \cite{GZ}, which is based on the demonstration fronts between subdomains undergo a random walk in the large size limit \cite{wpre}.

Let us term $x_k$ the positions of the $n$ fronts with $1\le k\le n$ separating the subdomains of length $d_k$. The spatial average of the field $A$ is approximately given by
\begin{equation}\bar{A}=\frac{1}{L}\sum_{k=1}^{n}\delta (-1)^k \underbrace{(x_k-x_{k-1})}_{=d_k}\,,\label{defafront}\end{equation}
where $\delta$ is a random variable taking values $\pm1$ with probability $1/2$. Note that we still work with the spatial average to capture order or lack thereof. It accounts for the possibility of having either $A=1$ or $A=-1$ in the first subdomain $0\le x\le x_1$. It is independent of the fronts positions, density of fronts \emph{etc}. Working on this simplified formulation will give us the main properties of the system in that regime. In order to calculate the average $\langle \bar{A}\rangle$ and the variance $\langle (\bar{A}-\langle \bar{A}\rangle)^2\rangle$ we will need the averages, variances and correlations of the $x_k$. Note that this do not take into accounts variations of $A$ around $\pm1$ in each subdomain: this should can taken into account by additional corrections, but does not play a major role in the transition. These fluctuations typically lead to the $1/\beta$ term in the response function of the mean field analysis (Eq.~(\ref{cmf})).

We will proceed in the following manner: we will remind the probability of having $n$ fronts in the limit of a non interacting gas of fronts.  This approximation assumes that the noise is not so large that no domain $|A|=1$ exist, but large enough that there are several in the domain. The fronts between subdomains $A=1$ and $A=-1$ will be small compared to the subdomain size. The fronts will be non interacting: this means that there is some cost of potential (in our gradient model) only for creating the fronts. The probability of having a configuration of fronts position and number is then controlled purely by entropy effects, given by the number of configurations of front positions. We will perform all averages in a systematic manner. We will determine beforehand the probability of having $n$ fronts by summing over all front positions. We will then separate the sums and integral of averages in two: we will first sum over all front positions at fixed numbers of front, then sum over front numbers. Additional calculation will be performed to obtain simple equivalents for the response function in the large size limit.

\subsubsection{Front positions statistics}

We first work out in details the case of dirichlet boundary conditions for the field $A$, so that $A(0)=A(L)=1$. In that case, we do not have translation invariance, but the number of fronts is not constrained to be even. Let us term $\Delta V$ the potential cost of creating a front we will use integrals of the type
\begin{eqnarray}
 \widetilde{f(\{ x_k\})}= \int_{x_1=0}^L\int_{x_2=x_1}^L\ldots \int_{x_n=x_{n-1}}^L f(\{ x_k\})e^{(-n\beta \Delta V)}\prod_{k=1}^n{\rm d}x_k\,.
\end{eqnarray}
In this integral, $\beta$ represents the variance of the local noise felt by the field $A$. If the dynamics of $A$ are gradient, $\Delta V$ represents the potential cost of creating a front. More generally, $\Delta V$ can represent the pseudo potential cost of creating a front in non gradient dynamics at low noise and high $\beta$. In more general non gradient cases $\exp(-\beta \delta V)$ can be replaced by the probability of creating a front, this will not change anything in the analysis and the results.
This integral has been introduced to calculate the probability $\widetilde{ 1}$ of having $n$ fronts in the domain \cite{GZ}. It takes into account both the cost of having $n$ fronts with the exponential as well as the number of configurations of the indistinguishable fronts. In particular, this has been used to show that the most probable number of fronts in the domain is $n\simeq L\exp(-\beta \Delta V)$. With that frame work the average of a function $f$, provided that there are $n$ fronts, is then given by $\langle f(\{x_k\})\rangle=\widetilde{f(\{x_k\})}/\widetilde{1}$. This uses the conditional probability of having fronts at given positions $\exp(-n\beta \Delta V)/\widetilde{1}$, provided that there are $n$ fronts, and sums over all the front positions. This first gives us the average position of front $k$
\begin{equation}
\langle x_{k}\rangle=L\frac{k}{n+1}
\end{equation}
This tells us that the average domain length is $\langle d_k\rangle=L/(n+1)$, which is a very natural result. The variance of $\bar{A}$ will include variances of $d_k$ and correlations between different domain lengths $d_k$ and $d_p$. The first stage in calculating them is to calculate variances and correlations of front positions. This yields
\begin{equation}
\langle x_{k}^2\rangle-\langle x_{k}\rangle^2=\frac{L^2k(n-k+1)}{(n+1)^2(n+2)}\label{eqvar}\,,
\end{equation}
and
\begin{equation}
\langle x_{k}x_{p}\rangle-\langle x_{k}x_{p}\rangle=\frac{L^2p(n-k+1)}{(n+1)^2(n+2)}\,.\label{eqcorrel}
\end{equation}
Of course the two formula coincide for $k=p$. Note that the correlation result above distinguishes $k$ and $p$ and assumes that $k\ge p$. From this, we can calculate the variance of a domain length
\begin{eqnarray}
\langle d_k^2\rangle -\langle d_k\rangle^2=\frac{L^2n}{(n+1)^2(n+2)}\label{vard}\,.
\end{eqnarray}
Note that this is independent of the front position. Since $n$ is proportional to $L$, this quantity is of order $\mathcal{O}(1)$ in $L$ in the large size limit. The variance of the domain size is a purely entropic effect arising from the equally probable front positions. We can then calculate the correlations between subdomain length $d_k$ and $d_p$. In order to distinguish the two subdomains, we set $k-1\ge p$, for proper use of the result of equation.~(\ref{eqcorrel}). We then find that
\begin{eqnarray}
\langle (d_k -\langle d_k\rangle )(d_p-\langle d_p\rangle)\rangle=-\frac{L^2}{(n+1)^2(n+2)}\label{cord}\,.
\end{eqnarray}
The correlations are independent of the distance between two subdomains, as long as they are different. The scaling in $1/L$ in the large size limit mainly arises from the constraint that $\sum_{k=1}^nd_k=L$. The domain size fluctuations are constrained by the total size. They are anticorrelated: if one grows, the other should retract.

\subsubsection{Average and variance}\label{dir}

Armed with these results, we can now compute the average and variance of the field. We have
\begin{equation}
\langle \bar{A}\rangle_n=\frac{1}{L}\langle \delta \rangle\sum_{k=1}^n(-1)^k\langle d_k\rangle=0\,.
\end{equation}
Since all the domain have the same average length, the sum is equal to either $-L/(n+1)$ if $n$ is odd or $0$ if $n$ is even. Meanwhile the average of $\delta$ is $0$. This is not much changed from the mean field case. We then calculate the variance of the spatially averaged field. We have
\begin{equation}
\eqalign{
\sigma^2_n=\cr \langle (\bar{A}-\langle \bar{A}\rangle_n)^2\rangle=\frac{1}{L^2}\left(\sum_{k=1}^n \langle (d_k-\langle d_k\rangle)^2\rangle +2\sum_{k=1}^n\sum_{l>k}^n (-1)^{l-k} \langle(d_k -\langle d_k\rangle )(d_l-\langle d_l\rangle)\rangle \right)\,.}
\end{equation}
We separated the sum of the variances of domain lengths and the sum of correlations of domain lengths, since these two quantities are distinct and independent of the index. The alternating sums from $l>k$ to $n$ are either $0$ ($n-k$ even) or $-1$ ($n-k$ odd). The calculation of the sums yields
\begin{equation}\label{fieldvar}\eqalign{
\sigma^2_n= \frac{1}{L^2}\left( n \langle (d_k-\langle d_k\rangle)^2\rangle+2\left\lfloor \frac{n}{2}\right\rfloor \langle(d_k -\langle d_k\rangle )(d_l-\langle d_l\rangle)\rangle \right) \cr= \frac{1}{L^2}\left( \frac{n^2L^2}{(n+1)^2(n+2)}+2\left\lfloor \frac{n}{2} \right\rfloor\frac{L^2}{(n+1)^2(n+2)} \right)\,,}
\end{equation}
where $\lfloor \cdot \rfloor$ denotes the integer part (\emph{i.e.} floor) function. In the limit of large $L$, the variance of domain size have a contribution which is of order $\mathcal{O}(1/L)$ , while the correlations between domain length have a contribution of order $\mathcal{O}(1/L^2)$ and are thus negligible. If we reintroduce by hand the dependence of most probable number of fronts $n$ on $L$ and $\beta$, we have in the large size limit that $\sigma^2\simeq \exp(\beta \Delta V)/L$.

In order to show this result, we now average $\sigma^2_n=\langle \bar{A}^2\rangle_n$ over the number of front. Our purpose is to obtain an equivalent for $\sigma$ in the large size limit so that we will either entirely calculate terms or give them bounds that ensure that they go to zero fast enough. Each has a probability $\widetilde{1}=(L\exp(-\beta \Delta V))^n/n!$ with $n\ge 1$ \cite{GZ}. Let us set $z=L\exp(-\beta \Delta V)$ to shorten the calculations. We recognise the normalisation $\sum_{n=1}^\infty z^n/n!=\exp(z)-1$. The variance reads
\begin{equation}
\sigma_2=\frac{1}{\exp(z)-1}\left(\underbrace{\sum_{n=1}^\infty \frac{n^2z^n}{(n+1)^2(n+2)n!}}_{=\Sigma_1}+\underbrace{\sum_{n=1}^\infty2\left\lfloor \frac{n}{2} \right\rfloor\frac{z^n}{(n+1)^2(n+2)n!}}_{=\Sigma_2} \right)\,.
\end{equation}
Let us first calculate $\Sigma_1$. We use operations on the sums of the type
\begin{equation}
\sum_{n=1}^\infty n a_n z^n=z\frac{d~}{dz}\sum_{n=1}a_n z^n\,,\,  \sum_{n=1}\frac{a_n}{n+1}z^n=\frac{1}{z}\int_{z'=0}^n\sum_{n=1}a_n{z'}^n\,{\rm dz'}\label{op}\,,
\end{equation}
to obtain
\begin{equation}
\Sigma_1=z\frac{d~}{dz}\left(z\frac{d~}{dz}\left(\frac{1}{z}\int_{z'=0}^z \left(\frac{1}{{z'}^2}\left(\int_{z''=0}^{z'}\left(\int_{z'''=0}^{z''}\left(\sum_{n=1}^\infty \frac{ {z'''}^n}{n!}\right)\,{\rm d}z'''\right){\rm d}z''\right)\right)\,{\rm d}z' \right)\right)\,.
\end{equation}
We recognise $\exp(z)-1$ in the sum and proceed with integrals and derivatives. We find that
\begin{equation}
\Sigma_1=\frac{\exp(z)-1-z}{z}-\frac{\exp(z)-1-z-\frac{z^2}{2}}{z^2} +\frac{1}{z}\underbrace{\int_{z'=0}^z \frac{\exp(z')-1-z'-\frac{{z'}^2}{2}}{{z'}^2}\,{\rm d}z'}_{=I}\,.
\end{equation}
One can insert the series expansion of exponential to check that each of the fraction is defined at $z=0$ and is in fact $0$, in agreement with the original sum. The first and third terms are positive and the second one is negative because, for $z\ge 0$, $\exp(z)$ minus its truncated series $1+z$, $1+z+z^2/2$, is a positive remainder $\sum_{n=k>1}^\infty z^n/n! $. We now give an upper bound to the integral which will be relevant provided $z\ge 1$. Let us split it into the integral from $0$ to $2$ (a given constant, smaller than $1$) and the integral from $2$ to $z$. In this second integral, the integrand is smaller than $f(z')=\exp({z'})/{z'}^2$ since $z\ge 1\ge 0$. Since $f'=(\exp(z')/{z'}^2)(1-2/z')$, the integrand $f(z')$ is growing over the whole interval and is always smaller that than $f(z)$. This means that
\begin{equation}\eqalign{
I=\int_{z'=0}^2 \frac{\exp(z')-1-z'-\frac{{z'}^2}{2}}{{z'}^2}\,{\rm d}z'+\int_{z'=2}^z \frac{\exp(z')-1-z'-\frac{{z'}^2}{2}}{{z'}^2}\,{\rm d}z'\cr \le 1+\int_{z'=2}^z \exp(z)/z^2\,{\rm d}z'=1+\frac{\exp(z)}{z}\label{b1}\,.}
\end{equation}
We now give bounds for $\Sigma_2$. It is positive, and using the fact that $\lfloor n/2\rfloor \le (n+1)/2$, we find that
\begin{equation}
0\le \Sigma_2\le \sum_{n=1}^\infty \frac{z^n}{(n+1)(n+2)n!}\label{b2}\,.
\end{equation}
Using operations of the type of equation~(\ref{op}), one finds that the upper bound is
\begin{equation}
\frac{1}{z^2}\left(\int_{z'=0}^z\left(\int_{z''=0}^z \left(\sum_{n=1}^{\infty}\frac{{z''}^n}{n!}\right)\,{\rm d}z''\right)\,{\rm d}z'\right)=\frac{\exp(z)-1-z-\frac{z^2}{2}}{2z^2}\,.\label{b3}
\end{equation}
Let us now show that $\sigma^2$ is equivalent to $z$ at infinity. We have
\begin{equation}
z\sigma^2=\frac{1-\exp(-z)(1-z)}{1-\exp(-z)}-\frac{1-(1+z+\frac{z^2}{2})\exp(-z)}{z(1-\exp(-z))}+a(z)\,.
\end{equation}
Using the bounds equation~(\ref{b1}),~(\ref{b2}) and~(\ref{b3}), we have that
\begin{eqnarray}
0\le a(z)\le \frac{1-(1+z+\frac{z^2}{2})\exp(-z)}{2z(1-\exp(-z)))}+\frac{1+\exp(-z)}{z(1-\exp(-z))}\,.
\end{eqnarray}
This means that
\begin{equation}
\lim_{z\rightarrow \infty}z\sigma^2=1\,.
\end{equation}
So that we have an equivalent for the variance
\begin{equation}
\sigma \sim_{L\rightarrow \infty} \sqrt{\exp(\beta \Delta V)/L}\,.
\label{eqdir}
\end{equation}

\subsubsection{periodic boundary conditions}\label{per}

The case of periodic boundary conditions can be considered in the same manner. We include the particularities of periodic boundary conditions.
Fronts appear by pairs, so that we only consider $n=2p$ fronts. And of course the domain is periodic, so that the first front is between $0$ and $L$, while the other ones are between $x_1$ and $x_1+L$, they still occupy a domain of size $L$, whose origin is shifted by $x_1$. We will therefore consider averages, conditioned to having $2p$ fronts, with integrals of the type
\begin{equation}
\widetilde{f\{x_k\}}=
 \int_{x_1=0}^L{\rm d}x_1\int_{x_2=x_1}^{L+x_1}{\rm d}x_2\ldots\int_{x_{2p}=x_{2p-1}}^{L+x_1}{\rm d}x_{2p}\, f(\{x_k\})e^{-2p\beta \Delta V}\,.
\end{equation}
The probability of having $2p$ fronts is thus $\Pi_{2p}=\widetilde{1}=\exp(-2p\beta \Delta V)L^{2p}/(2p-1)!$. One can show that the most probable number of par of fronts goes like $p\simeq L\exp(-\beta \Delta V)$. The average position of a front $k$ is then $\langle x_k\rangle=\frac{L(p+k-1)}{2p}$. The moments $\langle x_k^2\rangle$ and $\langle x_k x_q\rangle$ become quite involved. Careful rewriting yields a simple result for front position variance and correlations
\begin{eqnarray}
\langle x_k^2\rangle-\langle x_k\rangle^2=\frac{L^2}{12}+\frac{L^2(k-1)(2p-(k-1))}{(2p)(2p)(2p+1)}\,,
\\
\langle x_kx_q\rangle-\langle x_k\rangle \langle x_q\rangle=\frac{L^2}{12}+\frac{L^2(q-1)(2p-k+1)}{(2p)(2p)(2p+1)}\,.
\end{eqnarray}
There is one notable difference in structure for these cumulants, compared to the Dirichlet boundary conditions case (Eq.~(\ref{eqvar}), Eq.~\ref{eqcorrel}). The additive $L^2/12$ term comes from fluctuations of the whole front system as a block over a size $L$ allowed by the periodic boundary conditions. Note that $1/12$ is actually the variance of a homogeneous distribution on $[0;1]$. We then obtain the subdomain size fluctuations and correlations
\begin{equation}
\langle d_k^2\rangle-\langle d_k\rangle^2=\frac{L^2(2p-1)}{(2p)(2p)(2p+1)}\,, \, \langle (d_k-\langle d_k\rangle)(d_q-\langle d_q\rangle) \rangle=-\frac{L^2}{(2p)(2p)(2p+1)}\,.\label{vcp}
\end{equation}
We find similar scaling as in the Dirichlet boundary conditions case (Eq.~\ref{vard}, Eq.~\ref{cord}). Subdomain sizes are again anti-correlated. The average of the field is again $0$, and the variance of the field is unchanged when expressed with subdomain sizes variance and correlations (Eq.~\ref{fieldvar}). The floor $\lfloor n/2\rfloor=\lfloor 2p/2\rfloor=p$ is simpler. Inserting the values of equation~(\ref{vcp}) yields
\begin{equation}
\langle A^2\rangle|_{2p}-\langle A\rangle|_{2p}^2=\frac{(2p-1)}{(2p+1)(2p)}+\frac{1}{(2p)(2p+1)}\\=\frac{1}{2p+1}\,.
\end{equation}

In order to average over the front positions, we follow the same procedure using the weight $\Pi_{2p}$. Again, we set $z=L\exp(-\beta \Delta V)$. In all the following calculations, we will use the sum
\begin{eqnarray}
s(z)\equiv \sum_{z=1}^{\infty}z^{2p-1}/(2p-1)!=\sinh(z)\,.
\end{eqnarray}
One way to see this result is to insert the full power series of $\exp(\pm z)$ in the hyperbolic sine.
The normalisation reads $\sum{p=1}^\infty z^{2p}/(2p-1)!= z\sinh(z)$. The variance is thus
\begin{equation}
\sigma^2=\frac{\sum_{z=1}^{\infty}\frac{z^{2p-1}}{(2p+1)(2p-1)!}}{z\sinh(z)}=\frac{\frac{1}{z}\int_{z'=0}^z\left(z \sum_{z=1}^{\infty}\frac{z^{2p-1}}{(2p-1)!} \right)\,{\rm d}z'}{z\sinh(z)}
\end{equation}
The rewriting as an integral followed the lines of equation~(\ref{op}). After integration and simplification, we can show that we have an equivalent
\begin{equation}
z\sigma^2=\frac{1+\exp(-2z)-\frac{1}{z}-\frac{\exp(-2z)+2\exp(-z)}{z}}{1-\exp(-2z)}\,.
\end{equation}
We find again that $\lim_{z\rightarrow \infty} z\sigma^2=1$, so that we have the same equivalent $\sigma\sim_{L\rightarrow \infty}\sqrt{\exp(\beta \Delta V)/L}$ as in the Dirichlet boundary conditions case (Eq.~\ref{eqdir}). We can compare these two equivalent to the numerical simulations results. We first notice that in both simulations and analytics, we find that the spatial average of the order parameter has a monomodal distribution around zero in the large size limit. We also find that the response function $\sqrt{L}\sigma$ converges toward a size independent plateau in this large size limit. The convergence toward this plateau is slower and slower as $\beta$ is increased. Indeed, one first need a system large enough so that it can accommodate two domains of opposite sign of $A$ for the formalism to be relevant. Moreover, the equivalent is reached with a fixed rate in $z=L/\exp(\beta \Delta V)$. This means that if one wants $\sigma^2$ to be at a given distance from $1$, increasing $\beta$ means that $L$ must be increased correspondingly. We find that the amplitude of the large size limit value of $\chi$ increases exponentially with $\beta$, since $\chi\sim \exp(\beta \Delta V/2)$, in agreement with numerical simulations (Figure~\ref{simusAC} (f)). This $\beta$ effect means that the smearing of the transition is relevant for conceivable system sizes only if the product $\beta \Delta V$, cost of a front divided by noise variance, is not two large.

\subsection{Two dimensional case: Arguments for the scalings of the response function}

We first treated the one dimensional case, since it could be solved entirely analytically and extensive numerical simulations to check the scalings are affordable. However, even if this first study can pertain to crossover occurring wall flows extending in one dimension of space, it does not entirely enlighten us on the case of two dimensional systems. The statistics of front positions are not as directly derived as in the one dimensional case. Indeed, when we go from one to two dimensions, we go from point positions to a wide range of defect type (see \cite{CH} for a zoology). In our cases, the relevant defects are grain boundaries. In the simplest models and cases, the possibility of having grain boundaries and their type are controlled by the model type, its boundary conditions \emph{etc}. In our case the  large scale statistics appear to be well described by (close to) equilibrium statistical physics, as shown by the numerical results. However, the inner details of the model are actually very far from equilibrium. This leads to a very specific coarse grained model describing the state of the flow, for a large scale field $A$ taking distinct values where one finds banded turbulence or uniform turbulence. Indeed,  one finds only a finite number of directions for the grain boundaries (along $\vec e_x$ lines and along diagonal lines parallel to the band orientations).

The results obtained in the one dimensional case can give us hints for the explanation of the scalings of the response function. For this matter, we will directly base our argument on the variance and correlations of domain surfaces. We divide the total surface $S$ into the subdomains $S_i$ where $A$ takes the values $\pm 1$. There $n$ subdomains, such that $\sum_{i=1}^n S_i=S$. For a given realisation of $n$ and the $S_i$, the spatial average of the field $\bar{A}$ is
\begin{equation}
\bar{A}=\frac{1}{S}\delta\sum_{i}(-1)^i S_i\,.
\end{equation}
The ordering of the $S_i$ is such that the value of  $A$ in each subdomain is  $\delta (-1)^i$ and $\delta$ is the same random variable as in the former section, which takes value $\pm1$. We of course again have that $\langle \bar{A}\rangle=0$ so that the variance is again $\sigma^2=\langle \bar{A}^2\rangle$. Similarly to the one dimensional case, if one has $n$ subdomains the variance reads
\begin{equation}
\sigma^2_n=\frac{1}{S^2}\left(\sum_{i=1}^N \langle (S_i -\langle S_i)^2\rangle+2\sum_{l=1}^N\sum_{k>l} (-1)^{l-k}\langle (S_l-\langle S_l\rangle)(S_k-\langle S_k\rangle) \rangle  \right)\,.
\label{var2d0}
\end{equation}
One then average over $n$ to obtain the actual variance.
In order to calculate $\sigma$, we should perform the same series of computation, which yield the statistics of the $S_i$, then the ensemble average of $A$ at fixed $n$, then the average over $n$. Due to the variety of situations (grain boundary form \emph{etc}.), this task is far more technical than the derivation in one dimension. We can however argue that the same physical mechanisms are at play and that they will lead to very similar results. Indeed, the formation of this sustained density of grain boundaries leading to the domains is caused by a balance between the local noise (entropy) and the cost of creating a defect. This gives us the distribution of number of domains, domain sizes \emph{etc}. In particular, the most likely $n$ should scale like $S$, with a weigh coming from the probability of creating a grain boundary. Again, this should lead to a variance of each subdomain size of order $1$: one only finds some local fluctuations of the edges. We should also find correlation between domain sizes that go like $-1/S$. Indeed, the subdomain sizes are anti correlated, if one domain grows, some other shrinks. Moreover, the effect decreases with the total size: the larger the total domain, the less one subdomain is affect by a size change of another subdomain. In equation~(\ref{var2d0}), this means that the sum over the subdomain variances will lead to a term of order $n$. In the double sum over the subdomain sizes correlations, there are again global cancelations of alternating terms in the sum over $k$, so that one is left with the sum over $l$, this yields a term of order $n/S$, which will be small compared to the $n$ coming from the variances. As a consequence, one finds that $\sigma_n^2$ is of order $n/S^2$, that is to say of order $S$ with a prefactor coming from the probability of creating a grain boundary, so that $\sigma \propto 1/\sqrt{S}$. As a consequence, this means that the response function $\chi=\sqrt{S}\sigma$ should again be independent on size.

\end{document}